\documentclass[conference]{IEEEtran}
\IEEEoverridecommandlockouts

\ifCLASSINFOpdf
\usepackage[pdftex]{graphicx}
\graphicspath{ {images/} }
\else

\fi

\usepackage{cite}
\usepackage{amsmath,amssymb,amsfonts}
\usepackage{graphicx}
\usepackage{textcomp}
\usepackage{xcolor}
\usepackage{amsmath}
\usepackage{amsfonts}
\usepackage{amssymb,mathtools}
\usepackage{algpseudocode}
\usepackage[linesnumbered,ruled,vlined]{algorithm2e}
\usepackage{nomencl}
\def\BibTeX{{\rm B\kern-.05em{\sc i\kern-.025em b}\kern-.08em
    T\kern-.1667em\lower.7ex\hbox{E}\kern-.125emX}}

\begin{document}

\title{Multi-Layered Recursive Least Squares for Time-Varying System Identification}

\author {Mohammad Towliat, Zheng Guo, Leonard J. Cimini, Xiang-Gen Xia, and Aijun Song

\thanks{M. Towliat, L. J. Cimini and X.-G. Xia are with the Department of Electrical
and Computer Engineering, University of Delaware, Newark, DE, USA (e-mail:
mtowliat, cimini, xianggen@udel.edu)}
\thanks{Z. Guo and A. Song are with the Department of Electrical
and Computer Engineering, University of Alabama, Tuscaloosa, AL, USA (e-mail:
zguo18@crimson.ua.edu, song@eng.ua.edu)}
}

\maketitle
\begin{abstract}
Traditional recursive least square (RLS) adaptive filtering is widely used to estimate the  impulse responses (IR) of an unknown system. Nevertheless, the RLS estimator shows poor performance when tracking rapidly time-varying systems. In this paper, we propose a multi-layered RLS (m-RLS) estimator to address this concern. The m-RLS estimator is composed of multiple RLS estimators, each of which is employed to estimate and eliminate the misadjustment of the previous layer. It is shown that the mean square error (MSE) of the m-RLS estimate can be minimized by selecting the optimum number of layers. We  provide a method to determine the optimum number of layers. A low-complexity implementation of m-RLS is discussed and it is indicated that the complexity order of the  proposed estimator can be reduced to ${\cal O}(M)$, where $M$ is the IR length. In addition, by performing simulations, we show that m-RLS outperforms the classic RLS and the RLS methods with a variable forgetting factor.
\end{abstract}

\begin{IEEEkeywords}
Echo cancellation, mean square error, recursive least squares, system identification, time-varying systems
\end{IEEEkeywords}

%\vspace{-10pt}
\section{Introduction}

Recursive least squares (RLS) adaptive filtering is one of the most appealing frameworks for applications, such as system identification, inverse modeling, channel equalization, etc. \cite{R2}. The RLS algorithm, by using the Newton search, leads to a faster convergence than the steepest-decent based algorithms, e.g. least means squares (LMS) \cite{R7}. It has been shown that low-complexity versions of the RLS algorithm, including the fast transversal RLS (FTRLS) \cite{R11, R14, R15}, and the RLS lattice (RLSL) \cite{R6, R16} algorithms have almost the same convergence ability as the classic RLS, when tracking a time-varying system.
It is noteworthy that the adaptability of RLS filtering is based on the fact that the estimation error is a zero-mean white Gaussian variable and, thus,  the algorithm  recursively minimizes a maximum likelihood objective function. Accordingly, a regularized objective function can be introduced to preserve the tracking ability of RLS in a non-Gaussian environment \cite{R20, R21, R22, R23}.
 
In system identification, RLS is used to estimate the impulse response (IR) of an unknown system, given the input and the output of the system as the desired signal. Speaking of the convergence behavior of the RLS system identifiers, it has been shown in \cite{R9} that the estimation error is caused by two types of factors, including the lag error (LE) and the estimation noise (EN). The LE is due to the time-variation of the system. Since the RLS estimation is based on the statistical averages, it inherits a latency when tracking a time-varying IR. Thus, the LE can be problematic when estimating a fast time-varying system. On the other hand, the EN is caused by the exponentially windowing nature of RLS. Using a forgetting factor limits the actual observation window size and, even under steady-state conditions, the IR estimate is contaminated by a misadjustment.

Regarding the forgetting factor, there is a trade-off between the LE and the EN. A small forgetting factor raises the RLS tracking agility (i.e., a lower LE) but also develops a higher steady-state misadjustment (i.e., a higher EN). Contrarily, a larger forgetting factor leads to poor adaptability, but a better steady-state performance \cite{R10}. As a result, numerous studies in this field are dedicated to infer an RLS algorithm with a variable forgetting factor (RLS-VFF). For instance, see \cite{R3, R5, R26, R28}. The main idea in these works is to recognize the changes of the system's IR by measuring the changes in the desired signal statistics, then setting a small forgetting factor during the IR changes, and a large forgetting factor during the steady-state periods. 
 
In this paper, we look at the problem of low error system identification from another point of view. To this end, we propose the multi-layered RLS (m-RLS) algorithm that minimizes the sum of the LE and the EN.
The m-RLS approach is composed of multiple connected RLS estimators. Since the a posteriori error at each RLS is not completely uncorrelated with the input signal, the next RLS is employed to estimate and eliminate the error of the previous RLS. To this purpose, the a posteriori error of the previous RLS is considered as the desired signal in the next RLS. The total estimation error (sum of the LE and the EN) is a function of the number of layers. Therefore, by using the optimum number of layers, the estimation error of m-RLS can be less than that of RLS.
The optimum number of layers tightly depends on the coherence length of the system's IR (the time range in which the IR approximately remains invariant) and signal to noise ratio (SNR). We provide a solution for determining the optimal number of layers.

The complexity of m-RLS is discussed by comparing the number of multiplications, additions, and divisions of two types of implementations, including the classic and the transversal dichotomous coordinate descent (DCD) \cite{R30} techniques. We show that using transversal DCD implementation significantly reduces the complexity of m-RLS and makes the proposed method as a promising approach for the applications where superior adaptability is desired with the expense of a reasonable higher complexity. 

In our simulations, we evaluate the performance of m-RLS and show that, in a rapidly time-varying system, the proposed estimator leads to a lower estimation error compared to that of the classic RLS and the RLS-VFF techniques.

The rest of this paper is organized as follows. In Section II, we investigate the RLS system identification and the associated error when tracking a time-varying system. In Section III, we propose the m-RLS algorithm and provide a solution for the optimum number of layers that minimizes the estimation error of the proposed method. Section IV discusses the implementation and complexity of m-RLS. The simulation results are brought in Section V. Finally, Section VI concludes this paper. 

\textit{Notations}: Matrices are denoted by boldface uppercase
letters (e.g. $\mathbf{A}$), vectors are indicated by boldface lowercase letters (e.g. $\mathbf{a}$), and scalar quantities are presented by normal
letters (e.g. $a$ or $A$). ${{\mathbf{I}}_{M}}$ is the $M\times M$ identity matrix, and $\mathbb{E}$ is the mathematical expectation.  $\left\| \mathbf{a} \right\|$ indicates the $l_2$ norm of vector $\mathbf{a}$.
Finally, the superscripts ${{\left( . \right)}^{T}}$, ${{\left( . \right)}^{H}}$, and ${{\left( . \right)}^{*}}$ indicate transpose, conjugate transpose, and conjugate operators, respectively.

\section{RLS System Identification}
\subsection{RLS Algorithm}
Consider $x[n]$ as the input sequence to an unknown time-varying system with the IR vector ${\bf{h}}[n]={[{h_0}[n], \ldots ,{h_{M - 1}}[n]]^H}$, where $M$ is the length of IR.  The system's output is given as
\begin{equation}
d[n]={{{\bf{h}}^{H}[n]}}\mathbf{x}[n]+w[n],
\label{1}
\end{equation}
in which  $\mathbf{x}[n]={{\left[ x[n],\ldots ,x[n-M+1] \right]}^{T}}$ is the input signal vector, and $w[n]$ is an AWGN noise with variance $\sigma _{w}^{2}$. Given the input and the output signals, the system's IR can be estimated by using the well-known RLS algorithm as below \cite{R7}
\begin{subequations}
\begin{alignat}{4}
e[n] &= d[n] - {{{\bf{\hat h}}}^H}[n - 1]{\bf{x}}[n] \label{2_1}\\           
{\bf{k}}[n]{\rm{ }} &= {(\lambda  + {{\bf{x}}^H}[n]{\bf{P}}[n - 1]{\bf{x}}[n])^{ - 1}}{\bf{P}}[n - 1]{\bf{x}}[n] \label{2_2}\\
{\bf{\hat h}}[n] &= {\bf{\hat h}}[n - 1] + {e^*}[n]{\bf{k}}[n] \label{2_3}\\
{\bf{P}}[n] &= {\lambda ^{ - 1}}\left( {{\bf{I}}_{M} - {\bf{k}}[n]{{\bf{x}}^H}[n]} \right){\bf{P}}[n - 1], \label{2_4}
\end{alignat}
\end{subequations}
where ${\bf{\hat h}}[n]$  is the estimate of the IR, $d[n]$ is so called the desired signal, $0<\lambda < 1$ is the forgetting factor, $e[n]$ denotes the a priori error, and $\mathbf{k}[n]$ is the gain vector. It can be shown that in this algorithm, $\mathbf{k}[n]=\mathbf{P}[n]\mathbf{x}[n]$, where $\mathbf{P}[n]={{\mathbf{R}}^{-1}}[n]$, and $\mathbf{R}[n]=\sum\limits_{k=0}^{n}{{{\lambda }^{n-k}}}\mathbf{x}[k]{{\mathbf{x}}^{H}}[k]$ \cite{R2}. 

\subsection{Estimation Error in RLS}
In this section, we inquire about the accuracy of the RLS estimate, $\mathbf{\hat{h}}[n]$, by comparing it to the IR vector, ${\bf{h}}[n]$. For this evaluation, we use the mean square error (MSE) measure. By assuming that  $\mathbb{E}{\left\| {{\bf{h}}[n]} \right\|^2} = 1$, the normalized MSE is defined as
\begin{align}
\mu_{\text{RLS}} = \mathbb{E}{\left\| {{\bf{h}}[n] - {\bf{\hat h}}[n]} \right\|^2}.
\label{3_3}
\end{align} 
To facilitate analyzing $\mu_{\text{RLS}}$, we hold the following assumptions: 1) the input signal is an uncorrelated binary phase shift keying (BPSK) sequence, i.e., $x[n]\in \left\{ \pm 1 \right\}$; 2) the IR estimate is independent of the input signal \cite{R9}.

Since $\mathbb{E}\{\mathbf{x}[n]{{\mathbf{x}}^{H}}[n]\}=\mathbf{I}_M$, for a sufficiently large $n$, we have $\mathbf{R}[n]=\sum\limits_{k=0}^{n}{{{\lambda }^{n-k}}}=\frac{1}{\varepsilon }\mathbf{I}_M$, where $\varepsilon =  1-\lambda $. As a result, ${\bf{P}}[n] = \varepsilon {\bf{I}}_M$ and ${\bf{k}}[n] = \varepsilon {\bf{x}}[n]$.
On the other hand, by replacing \eqref{1} in \eqref{2_1}, the a priori error is given as $e[n] = {( {{\bf{ h}}[n] - {\bf{\hat h}}[n - 1]})^H}{\bf{x}}[n] + w[n]$.
Nesting ${\bf{k}}[n]$ and $e[n]$ in \eqref{2_3}, the IR estimate at time $n$th becomes
\begin{align}
{\bf{\hat h}}[n] &= {\bf{h}}[n] - {\bf{\Theta }}[n]\left( {{\bf{h}}[n] - {\bf{\hat h}}[n - 1]} \right) + {\boldsymbol{\gamma }}[n],
\label{5}
\end{align} 
where $\mathbf{\Theta }[n]={{\mathbf{I}}_{M}}-\varepsilon \mathbf{x}[n]{{\mathbf{x}}^{H}}[n]$ and ${\boldsymbol{\gamma }}[n]=\varepsilon {{w}^{*}}[n]\mathbf{x}[n]$.  Similar to \eqref{5}, the IR estimate at the past times can be written as $\mathbf{\hat{h}}[n-k]={\bf{h}}[n-k]-\mathbf{\Theta }[n-k] ( {\bf{h}}[n-k]-\mathbf{\hat{h}}[n-k-1])+{\boldsymbol{\gamma }}[n-k]$. By using this recursion, for $k = 0, \ldots ,N-1$, one can expand \eqref{5} as
\begin{equation}
{\bf{\hat h}}[n] = {\bf{A}}[n]{\bf{h}}[n] + {\bf{B}}[n]{\bf{\hat h}}[n - N] + {\bf{c}}[n],
\label{7}
\end{equation}
where 
\begin{align}
{\bf{A}}[n] &= \varepsilon {\bf{x}}[n]{{\bf{x}}^H}[n] \nonumber\\
&\,\,\,\,\,\,+ \varepsilon \sum\limits_{k = 1}^{N - 1} {\left( {\prod\limits_{i = 0}^{k - 1} {{\bf{\Theta }}[n - i]} } \right){\bf{x}}[n - k]{{\bf{x}}^H}[n - k]};  \nonumber\\
 {\bf{B}}[n] &= \prod\limits_{k = 0}^{N - 1} {{\bf{\Theta }}[n - k]};  \nonumber\\
{\bf{c}}[n] &= {\boldsymbol{\gamma }}[n] + \sum\limits_{k = 1}^{N - 1} {\left( {\prod\limits_{i = 0}^{k - 1} {{\bf{\Theta }}[n - i]} } \right)} {\boldsymbol{\gamma }}[n - k],
\label{7_2}
\end{align} 
and $N$ is the coherence length of the IR (the time range for which the IR remains  invariant), so that ${\bf{h}}[n-k] = {\bf{h}}[n]$, for $k = 0, \ldots ,N - 1$. Note that, for a relatively slow time-varying IR, $N$ is large while, for a fast time-varying IR, $N$ can be small. 

By doing some algebra, it can be shown that ${{\bf{I}}_M} - {\bf{A}}[n] = {\bf{B}}[n]$. As a result, the difference between the true IR and its RLS estimate in \eqref{7} becomes
\begin{align}
{\bf{h}}[n] - {\bf{\hat h}}[n]={\bf{B}}[n]({\bf{h}}[n] - {\bf{\hat h}}[n - N]) - {\bf{c}}[n]. 
\label{77_2}
\end{align} 
On the right-hand side of  \eqref{77_2}, ${\bf{c}}[n]$ is the effect of the noise samples $w[n-k]$ for $k = 0, \ldots ,N - 1$, while ${\bf{\hat h}}[n - N]$  depends on $w[n-k]$ for $k \ge N$. Since ${\bf{c}}[n]$ is uncorrelated with both ${\bf{\hat h}}[n - N]$ and ${\bf{h}}[n]$,  replacing \eqref{77_2} in \eqref{3_3} leads to
\begin{align}
\mu_{\text{RLS}}  =& \mathbb{E}{\left\| {\bf{B}}[n]({\bf{h}}[n] - {\bf{\hat h}}[n - N]) \right\|^2} + \mathbb{E}{\left\| {{{\bf{ c}}}[n]} \right\|^2}.
\label{77_4}
\end{align}
We have shown in Appendix A that \eqref{77_4} can be simplified to
\begin{align}
\mu_{\text{RLS}} =  & {\rho ^N}\mathbb{E}{\left\| {{\bf{h}}[n]} \right\|^2} + \left( {1 - {\rho ^{N + 1}}} \right)\psi \sigma _w^2 \nonumber\\
=&{\rho ^N}+\left( {1 - {\rho ^{N + 1}}} \right)\psi \sigma _w^2,  
\label{77_5}
\end{align}
where $\rho  = 1 - 2\varepsilon  + {\varepsilon ^2}M$, and $\psi  = \frac{{{\varepsilon ^2}M}}{{1 - \rho }}$. Note that, when the IR is time-invariant, Eq. \eqref{77_5} coincides with the known result \cite{R2} and \cite{R9}.

According to \eqref{77_5}, the RLS estimation error is contributed by two terms \cite{R9, R13}. The first term is the LE caused by the RLS lag of tracking a time-varying IR. The second term is the EN, which is the effect of exponentially windowing nature of RLS, so that by letting $\lambda  = 1$, this term is zero. 

Speaking of the LE, we should mention that it originates from the fact that the RLS estimation is performed based on statistical averages which take several time samples to  converge. In a time-varying system, the averaging process can be inadequate because of  the rapid variations. 
As mentioned in \cite{R11} and \cite{R10}, for an stable convergence, the forgetting factor falls within the range $1 - \frac{2}{M} < \lambda  < 1$. In this range, one can see that $0< \rho < 1$.
Accordingly, when the IR is relatively slow time-varying with a large $N$, the contribution of the LE in \eqref{77_5} is small and the MSE is mainly governed by the EN. On the other hand, for a rapidly time-varying IR with a small $N$, the LE is not negligible and can potentially be higher than that of the EN.

Addressing the problem of minimizing the MSE in tracking a rapidly time-varying system, we propose the m-RLS algorithm in the next section.

\section{Multi-Layered RLS System Identification}
Concerning about low estimation error in tracking time-varying systems, we propose the m-RLS estimator to minimize the sum of LE and EN. The proposed m-RLS estimator is composed of multiple layers. At each layer, the effective IR (which is the estimation error of the previous layer) is estimated and eliminated from the desired signal by utilizing a separate RLS estimator.
We show that, by using the optimum number of layers, the m-RLS strategy leads to a lower MSE error than RLS.

\subsection{Multi-layered RLS Algorithm}
\begin{figure*}[bt!]
  \includegraphics[width=6in]{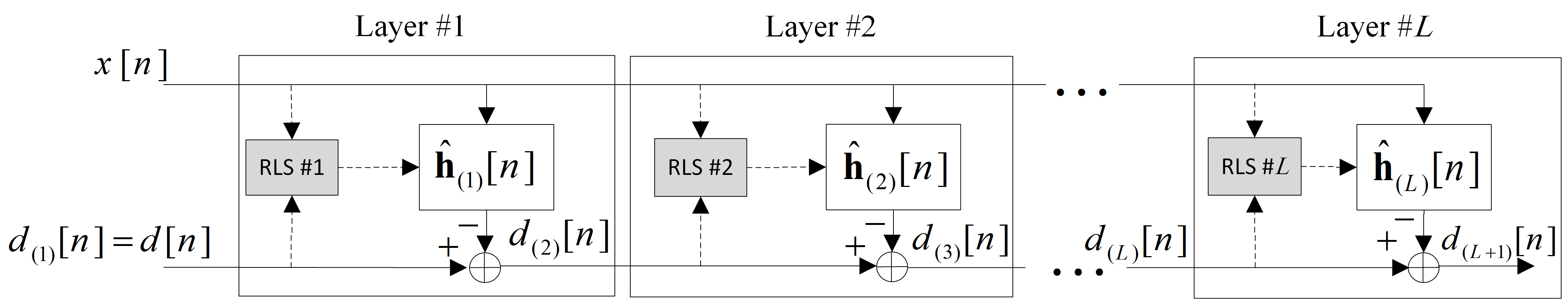}
  \caption{The m-RLS system identifier structure.}
	\vspace{-10pt}
	\label{F1}
\end{figure*}

Fig.~\ref{F1} shows the m-RLS structure with $L$ layers. Similar to the RLS estimator, m-RLS is used to estimate the IR given $d[n]$ and ${\bf x}[n]$. 

In the first layer, the desired signal is $d[n]$, thus, we  represent it with ${{d}_{(1)}}[n]=d[n]$, where the subscription denotes the layer's index. Accordingly, we can rewrite \eqref{1} as 
\begin{equation}
{d_{(1)}}[n] = {\bf{h}}_{(1)}^H[n]{\bf{x}}[n] + w[n],
\label{8}
\end{equation}
where ${{\mathbf{h}}_{(1)}}[n]={{\mathbf{h}}}[n]$  is the effective IR at the first layer. RLS \#1  estimates ${{\mathbf{h}}_{(1)}}[n]$ given ${d_{(1)}}[n]$ and ${\bf{x}}[n]$. We denote this estimate as ${{\bf{\hat h}}_{(1)}}[n]$.  
Then, the a posteriori error \cite{R2} in RLS \#1 is given as
\begin{align}
{d_{(2)}}[n] &= {d_{(1)}}[n] - {\bf{\hat h}}_{(1)}^H[n]{\bf{x}}[n] \nonumber\\
 &= {\left( {{{\bf{h}}_{(1)}}[n] - {{{\bf{\hat h}}}_{(1)}}[n]} \right)^H}{\bf{x}}[n] + w[n] \nonumber\\
 &= {\bf{h}}_{(2)}^H[n]{\bf{x}}[n] + w[n].
\label{10}
\end{align} 
where ${{\bf{h}}_{(2)}}[n]= {{\bf{h}}_{(1)}}[n] - {{\bf{\hat h}}_{(1)}}[n]$, is the error vector of estimating ${{\mathbf{{h}}}_{(1)}}[n]$ by using RLS \#1. Due to the LE of RLS \#1, ${{\mathbf{{h}}}_{(2)}}[n]$ is not a pure noise vector and, thus, ${d_{(2)}}[n]$ has a correlation with the input signal ${\bf{x}}[n]$. 

At the second layer, RLS \#2 is employed to provide an estimate of the effective IR ${{\mathbf{h}}_{(2)}}[n]$. To this end,  ${d_{(2)}}[n]$ is taken as the desired signal. We denote the estimate of ${{\bf{h}}_{(2)}}[n]$ by ${{\bf{\hat h}}_{(2)}}[n]$. Similar to the first layer, the a posteriori error in RLS \#2 becomes
\begin{align}
 {d_{(3)}}[n] &= {d_{(2)}}[n] - {\bf{\hat h}}_{(2)}^H[n]{\bf{x}}[n] \nonumber\\
&= {\bf{h}}_{(3)}^H[n]{\bf{x}}[n] + w[n], 
\label{10_}
\end{align} 
where ${{\bf{h}}_{(3)}}[n] = {{\bf{h}}_{(2)}}[n] - {{{\bf{\hat h}}}_{(2)}}[n]$. 

The same process is performed for all layers such that, at the $l$th layer (for $l = 1, \ldots ,L$), the a posteriori error is 
\begin{align}
{d_{(l+1)}}[n] &= {d_{(l)}}[n] - {\bf{\hat h}}_{(l)}^H[n]{\bf{x}}[n] \nonumber\\
 &  = {\bf{h}}_{(l+1)}^H[n]{\bf{x}}[n] + w[n],
\label{12}
\end{align} 
where ${d_{(l)}}[n] = {\bf{h}}_{(l)}^H[n]{\bf{x}}[n] + w[n]$, and
\begin{equation}
{{\mathbf{h}}_{(l+1)}}[n]={{\mathbf{h}}_{(l)}}[n]-{{\mathbf{\hat{h}}}_{(l)}}[n].
\label{12_add}
\end{equation}
Finally, the overall IR estimation by m-RLS is given as the sum of the estimates at all layers, that is 
\begin{equation}
{\bf{\tilde h}}[n] = \sum\limits_{l = 1}^L {{{{\bf{\hat h}}}_{(l)}}[n]}.
\label{13}
\end{equation}

%\vspace{-10pt}
\subsection{Estimation Error in Multi-Layered RLS}

In this section, we evaluate the accuracy of the proposed m-RLS estimator. To facilitate the derivations, we hold the same assumptions as those in Section II-B.

The overall IR estimate by m-RLS is given in \eqref{13}. The normalized MSE of this estimate is obtained as
\begin{align}
{\mu _{{\rm{m \textendash RLS}}}}(L) = \mathbb{E}{\left\| {{\bf{h}}[n] - {\bf{\tilde h}}[n] } \right\|^2}.
\label{13_1}
\end{align} 
In \eqref{13_1}, we represent the MSE as a function of $L$ because, according to \eqref{13}, ${\bf{\tilde h}}[n]$ is introduced based on the number of layers, $L$.
By considering the recursions in \eqref{12_add}, substituting \eqref{13} in \eqref{13_1} results in
\begin{align}
{\mu _{{\rm{m \textendash RLS}}}}(L) = \mathbb{E}{\left\| {{{\bf{h}}_{(L + 1)}}[n]} \right\|^2}.
\label{13_2}
\end{align} 

Eq. \eqref{13_2} indicates that the MSE of m-RLS is equivalent to the average power of the last effective IR, $ {{{\bf{h}}_{(L + 1)}}[n]}$. Thus, let us first evaluate the general case $\mathbb{E}{\left\| {{{\bf{h}}_{(l + 1)}}[n]} \right\|^2}$, for $l = 1, \ldots L$.  In the proposed estimator, each RLS operates separately; therefore, similar to \eqref{77_2}, we can represent the difference between the effective IR and its estimate at the $l$th layer as
\begin{align}
{{\bf{h}}_{(l+1)}}[n]=&{{\bf{h}}_{(l)}}[n] - {{{\bf{\hat h}}}_{(l)}}[n] \nonumber\\
=& {{\bf{B}}_{(l)}}[n]({{\bf{h}}_{(l)}}[n] - {{{\bf{\hat h}}}_{(l)}}[n - N_{(l)}]) - {{\bf{c}}_{(l)}}[n],
\label{16_20}
\end{align}
where 
\begin{align}
 {\bf{B}}_{(l)}[n] &= \prod\limits_{k = 0}^{N_{(l)} - 1} {{\bf{\Theta }}[n - k]};  \nonumber\\
{{\bf{c}}_{(l)}}[n] &= {\boldsymbol{\gamma }}[n] + \sum\limits_{k = 1}^{{N_{(l)}} - 1} {\left( {\prod\limits_{i = 0}^{k - 1} {{\bf{\Theta }}[n - i]} } \right)} {\boldsymbol{\gamma }}[n - k],
\label{16_21}
\end{align} 
and $N_{(l)}$ is the coherence length of ${{\bf{h}}_{(l)}[n]}$.

In \eqref{16_20}, ${{{\bf{c}}_{(l)}}[n]}$ is based the noise samples $w[n-k]$ for $k = 0, \ldots ,N_{(l)} - 1$, whereas ${\bf{\hat h}}_{(l)}[n - N_{(l)}]$  depends on $w[n-k]$ for $k \ge N_{(l)}$. Thus, ${{{\bf{c}}_{(l)}}[n]}$  is uncorrelated with ${{{\bf{\hat h}}}_{(l)}}[n - {N_{(l)}}]$. On the other hand, ${{{\bf{c}}_{(l)}}[n]}$ has a correlation with ${{{\bf{h}}_{(l)}}[n]}$, except for $l=1$ (see \eqref{77_4}). The correlation between ${{{\bf{c}}_{(l)}}[n]}$ and ${{{\bf{h}}_{(l)}}[n]}$ for $l>1$, leads to
\begin{align}
\mathbb{E}{\left\| {{{\bf{h}}_{(l + 1)}}[n]} \right\|^2}=& \mathbb{E}{\left\| {{{\bf{B}}_{(l)}}[n]({{\bf{h}}_{(l)}}[n] - {{{\bf{\hat h}}}_{(l)}}[n - {N_{(l)}}])} \right\|^2} \nonumber\\
& + \mathbb{E}{\left\| {{{\bf{c}}_{(l)}}[n]} \right\|^2}+ u(l),
\label{16_2111}
\end{align}
where  $u(l)=- 2{\rm{Re}}\mathbb{E}\{ {{\bf{c}}_{(l)}^H[n]{{\bf{B}}_{(l)}}[n]{{\bf{h}}_{(l)}}[n]}\}$ is the cross-correlation term. 
Following the same steps as those in Appendix A, the average power of ${{\bf{h}}_{(l+1)}}[n]$  can be represented as
\begin{align}
\mathbb{E}{\left\| {{{\bf{h}}_{(l + 1)}}[n]} \right\|^2} = &{\rho ^{{N_{(l)}}}}\mathbb{E}{\left\| {{{\bf{h}}_{(l)}}[n]} \right\|^2} + \left( {1 - {\rho ^{{N_{(l)}} + 1}}}\right)\psi \sigma _w^2 \nonumber\\
& +u(l).
\label{16_23}
\end{align}

Eq. \eqref{16_23} exhibits the recursive relation between the average powers of two consecutive effective IR vectors ${{\bf{h}}_{(l+1)}}[n]$ and ${{\bf{h}}_{(l)}}[n]$, for $l = 1, \ldots ,L$.
By using this recursion, one can expand $\mathbb{E}{\left\| {{{\bf{h}}_{(L + 1)}}[n]} \right\|^2}$ based on $\mathbb{E}{\left\| {{{\bf{h}}_{(1)}}[n]} \right\|^2}=1$ and obtain ${\mu _{{\rm{m \textendash RLS}}}}(L)$ in \eqref{13_2} as
\begin{align}
&{\mu _{{\rm{m \textendash RLS}}}}(L) = \mathbb{E}{\left\| {{{\bf{h}}_{(L + 1)}}[n]} \right\|^2} = \prod\limits_{l = 1}^L {{\rho ^{{N_{(l)}}}}}  + v(L).
\label{16_25}
\end{align}
where $v(L) = {\rho ^{{N_{(L)}}}}v(L - 1) + \left( {1 - {\rho ^{{N_{(L)}} + 1}}} \right) \psi \sigma _w^2 + u(L)$ is a recursive function initiated as $v(1)={1 - {\rho ^{{N_{(1)}} + 1}}}$ and $u(1)=0$.

According to \eqref{16_25}, ${\mu _{{\rm{m \textendash RLS}}}}(L)$ is the sum of two terms. $\prod\limits_{l = 1}^L {{\rho ^{{N_{(l)}}}}}$ is the LE, originating from the system time-variations, and $v(L)$ is the EN, representing the noise effect.
The values of the LE and the EN tightly depend on the coherence lengths of the effective IRs, ${{N_{(l)}}}$, for $l = 1, \ldots ,L$. Thus, prior to discussing ${\mu _{{\rm{m \textendash RLS}}}}(L)$, let us first investigate ${{N_{(l)}}}$ in the next section.

\subsection{Coherence Length of Effective IRs}
In this section, we investigate the coherence lengths of the effective IRs when SNR is high.
Let ${\varphi _{(l + 1)}}[m]$, for $m \ge 0$, be the normalized autocorrelation function (ACF) of ${{{\bf{h}}_{(l+1)}}[n]}$ calculated as
\begin{align}
{\varphi _{(l + 1)}}[m] = \frac{\mathbb{E}\{{\bf{h}}_{(l + 1)}^H[n]{{\bf{h}}_{(l + 1)}}[n - m]\}}{{\mathbb{E}{{\left\| {{{\bf{h}}_{(l + 1)}}[n]} \right\|}^2}}}.
\label{17_2}
\end{align}
In \eqref{17_2}, it is assumed that all tap-weights in the effective IR vector share the same ACF. When the SNR is high (i.e., $\sigma _w^2$ is small and negligible), replacing \eqref{16_20} in \eqref{17_2} gives us (see Appendix B)
\begin{align}
&{\varphi _{(l + 1)}}[m] = \nonumber\\
&\left( {2{\varphi _{(l)}}[m] - {\varphi _{(l)}}[m - {N_{(l)}}] - {\varphi _{(l)}}[m + {N_{(l)}}]} \right) q_{(l)}[m],
\label{17_5}
\end{align}
where
\begin{align}
{q_{(l)}}[m] = \left\{ {\begin{array}{*{20}{l}}
{{{\left( {\frac{{{\lambda ^2}}}{\rho }} \right)}^m};{\kern 1pt} {\kern 1pt} {\kern 1pt} {\kern 1pt} {\rm{for}}{\kern 1pt} \,\,0 \le m \le {N_{(l)}}{\kern 1pt} }\\
{\kern 1pt} \\
{{{\left( {\frac{{{\lambda ^2}}}{\rho }} \right)}^{{N_{(l)}}}};{\kern 1pt} {\kern 1pt} {\kern 1pt} {\kern 1pt} {\rm{for}}\,\,{\kern 1pt} {N_{(l)}} < m,{\kern 1pt} {\kern 1pt} }
\end{array}} \right.
\label{17_6}
\end{align}
and ${\varphi _{(l)}}[m]$ is the normalized ACF of ${{{\bf{h}}_{(l)}}[n]}$.

Given the normalized ACF of a random variable, the coherence length of the variable is defined as the interval for which, the normalized ACF is greater than 0.5 \cite{R31}. By using this fact we show in Appendix C that, if  ${\varphi _{(l)}}[m]$ has an exponential shape as ${\varphi _{(l)}}[m] = \exp ( - \alpha m )$, with $\alpha  > 0$, then Eq. \eqref{17_5} leads to the conclusion that ${\varphi _{(l+1)}}[m]$ also inherits an exponential form expressed as ${\varphi _{(l + 1)}}[m] \approx \exp (-\beta  m )$, for $0 \le m \le {N_{(l)}}$, where $\beta = 3\alpha + g$ and  $g = \log (\frac{\rho }{{{\lambda ^2}}})$. 

By letting ${\varphi _{(l)}}[{N_{(l)}}]=0.5$ and ${\varphi _{(l+1)}}[{N_{(l+1)}}]=0.5$, the exponential shapes of ${\varphi _{(l+1)}}[m]$ and ${\varphi _{(l+1)}}[m]$ lead to ${N_{(l)}} = \frac{{\log 2}}{\alpha }$ and ${N_{(l + 1)}} \approx  \frac{{\log 2}}{\beta } = \frac{{\log 2}}{{3\alpha  + g}}$, respectively. Combining the last two achievements gives
\begin{align}
{N_{(l + 1)}} \approx \left\lceil {\frac{{{N_{(l)}}}\text{log}2}{{{N_{(l)}}g + 3\text{log}2}}} \right\rceil.
\label{17_666}
\end{align}
where $\left\lceil  \bullet  \right\rceil$ denotes the ceiling operator.

Eq. \eqref{17_666} highlights two important points about the coherence lengths. First, it shows ${N_{(l + 1)}} \le {N_{(l)}}$ indicating that ${{{\bf{h}}}_{(l+1)}}[n]$ is more time-varying than ${{{\bf{h}}}_{(l)}}[n]$. To physically explain this achievement, consider \eqref{12_add}, which demonstrates that the variations in ${{{\bf{h}}}_{(l+1)}}[n]$ is the superposition the variations in ${{{\bf{h}}}_{(l)}}[n]$ and ${{{\bf{\hat h}}}_{(l)}}[n]$. Because of the lag in ${{{\bf{\hat h}}}_{(l)}}[n]$, its variation can be considered noncoherent with that of ${{{\bf{ h}}}_{(l)}}[n]$ in a specific sample time. As a result, ${{{\bf{h}}}_{(l+1)}}[n]$ becomes faster fluctuating than  ${{{\bf{h}}}_{(l)}}[n]$.

The second point from \eqref{17_666} is that, by using this recursion, all coherence lengths can be approximately represented based on the coherence lengths of the effective IR at the first layer, ${N_{(1)}} = N$ (note that the effective IR at the first layer, ${{\bf{h}}_{(1)}}[n]$, is equivalent to the system's IR, ${{\bf{h}}}[n]$, with the coherence length $N$).

\subsection{Optimal Number of Layers}

Eq. \eqref{16_25} shows that both the LE and the EN in m-RLS are functions of the number of layers, $L$. The optimum number of layers, ${L_{{\rm{opt}}}}$, is where the sum of the LE and the EN (i.e., the total MSE in \eqref{16_25}) is minimized, that is
\begin{align}
{L_{{\rm{opt}}}} = \arg \min_{l} \quad & \mathbb{E}{\left\| {{{\bf{h}}_{(l + 1)}}[n]} \right\|^2} \nonumber\\ 
\textrm{s.t.} \quad & 1 \le l \le {L_{\max }},
\label{166_26}
\end{align}
where ${L_{\max }}$ is the maximum allowed number of layers to keep the complexity of m-RLS bounded. Note that by comparing \eqref{16_25} and \eqref{77_5}, it is seen that ${\mu _{{\rm{m \textendash RLS}}}}(1)={\mu _{{\rm{RLS}}}}$. This indicates that m-RLS with one layer is identical to an RLS estimator (this fact is also seen in Fig.~\ref{F1}).
With this respect, one can see that ${L_{{\rm{opt}}}}$ in \eqref{166_26} leads to
\begin{align}
{\mu _{{\rm{RLS}}}} \ge {\mu _{{\rm{m  \textendash RLS}}}}({L_{{\rm{opt}}}}).
\label{166_27}
\end{align}
According to \eqref{166_27}, by setting the number of layers to ${L_{{\rm{opt}}}}$, m-RLS has a more accurate estimate than RLS, if ${L_{{\rm{opt}}}}>1$. The equality holds only for the case where ${L_{{\rm{opt}}}}=1$.

As mentioned in Section III-C, $N_{(l)}$ in \eqref{16_25} can be introduced based on $N$. As a result, one can realize that ${L_{{\rm{opt}}}}$  in \eqref{166_26} only depends on two parameters, including the coherence length of the system's IR, $N$, and the noise power, $\sigma _w^2$. Later, in Section V, we infer that the solution of \eqref{166_26} ends to a larger ${L_{{\rm{opt}}}}$, for either a smaller $N$ or a smaller $\sigma _w^2$. Alternatively stated, for either a relatively faster time-varying systems or a higher SNR, ${L_{{\rm{opt}}}}$ becomes larger. 

Nevertheless, it is difficult to provide a closed-form solution for \eqref{166_26}.
Therefore, we use a numerical method to determine ${L_{{\rm{opt}}}}$.
To this purpose, in Appendix D, we show that $\mathbb{E}{\left\| {{{\bf{h}}_{(l + 1)}}[n]} \right\|^2}$ can be written based on the average power of the a posteriori error as
\begin{align}
\mathbb{E}{\left\| {{{\bf{h}}_{(l + 1)}}[n]} \right\|^2} = \mathbb{E}{\left| {{d_{(l + 1)}}[n]} \right|^2} +(1- 2{(1 - \varepsilon M)^l})\sigma _w^2.
\label{166_277}
\end{align}
Thus, the optimization problem in \eqref{166_26} is equivalent to the following minimization
\begin{align}
\begin{array}{l}
{L_{{\rm{opt}}}} = \mathop {\arg \min }\limits_l \,\,\,\mathbb{E}{\left| {{d_{(l + 1)}}[n]} \right|^2} - 2{(1 - \varepsilon M)^l}\sigma _w^2.\\
\,\,\,\,\,\,\,\,\,\,\,\,\,\,\,\,\,\,\, \text{s.t.}\,\,\, 1 \le l \le {L_{\max }}
\end{array}
\label{166_288}
\end{align}

In \eqref{166_288}, $\mathbb{E}{\left| {{d_{(l + 1)}}[n]} \right|^2}$ can be numerically estimated as 
\begin{align}
{\pi _{(l + 1)}}[n] =& (1 - z){\pi _{(l + 1)}}[n - 1] +z {\left| {{d_{(l + 1)}}[n]} \right|^2}, 
\label{19_1}
\end{align}
where ${\rm{0 < z}} \ll {\rm{1}}$ is a small scalar, and ${{\pi }_{(l+1)}}[n]$ is the real-time estimate of $\mathbb{E}{{\left| {{d}_{(l+1)}}[n] \right|}^{2}}$.

\section{Implementation and Complexity Analysis}

Each layer of m-RLS is equipped with a separate RLS estimator as described in \eqref{2_1}-\eqref{2_4}. However, since ${\bf{k}}[n]$ and ${\bf{P}}[n]$ only depend on the forgetting factor and the input signal, all layers in m-RLS share the same ${\bf{k}}[n]$ and ${\bf{P}}[n]$, which results in a simpler structure.

Similar to \eqref{2_1}, the a priori error of the RLS \#$l$ is given as ${e_{(l)}}[n] = {d_{(l)}}[n] - {\bf{\hat h}}_{(l)}^H[n - 1]{\bf{x}}[n]$ and, similar to \eqref{2_3}, the effective IR is updated as ${{{\bf{\hat h}}}_{(l)}}[n] = {{{\bf{\hat h}}}_{(l)}}[n - 1] + e_{(l)}^*[n]{\bf{k}}[n]$. 
After that, the a posteriori error is calculated as
\begin{align}
{d_{(l + 1)}}[n] =& {d_{(l)}}[n] - {\bf{\hat h}}_{(l)}^H[n]{\bf{x}}[n] \nonumber\\ 
 =& \left( {{d_{(l)}}[n] - {\bf{\hat h}}_{(l)}^H[n - 1]{\bf{x}}[n]} \right) \nonumber\\ 
&- {\left( {{{{\bf{\hat h}}}_{(l)}}[n] - {{{\bf{\hat h}}}_{(l)}}[n - 1]} \right)^H}{\bf{x}}[n]
= {e_{(l)}}[n] T[n],
\label{199_7}
\end{align}
where $T[n]={1 - {{\bf{k}}^H}[n]{\bf{x}}[n]}$.  As a result, 
the m-RLS algorithm can be summarized as shown in Algorithm 1, where $0<\delta \ll 1$ and $0<z \ll 1$ are arbitrary quantities. 
\begin{algorithm}
    \caption{m-RLS algorithm.}
    \DontPrintSemicolon
    \textbf{Initialization:} 
    \quad  ${\bf{P}}[ - 1] = {\delta ^{ - 1}}{{\bf{I}}_M}$, ${\bf{k}}[-1]=\bf{0}$, ${{{{\bf{\tilde h}}}}}[-1]=\bf{0}$,  ${{{{\bf{\hat h}}}_{(l)}}}[-1]=\bf{0}$,  ${{{{{\pi}}}_{(l+1)}}}[-1]={0}$, $r(l) = 2{(1 - \varepsilon M)^l}\sigma _w^2$, for $l = 1, \ldots ,{L_{{\rm{max}}}}$.

    \For  {$n = 1,2,3, \ldots$} {
					
					 ${\bf{k}}[n] = {\left( {\lambda + {{\bf{x}}^H}[n]{\bf{P}}[n - 1]{\bf{x}}[n]} \right)^{ - 1}}{\bf{P}}[n - 1]{\bf{x}}[n]$
			
			     ${{{\bf{P}}}[n]}={  \lambda^{ - 1}\left( {{\bf{I}}_M - {{\bf{k}}}[n]{{\bf{x}}^H}[n]} \right){{\bf{P}}}[n - 1]}$
			
			     $T[n] = 1 - {{\bf{k}}^H}[n]{\bf{x}}[n]$
					
					$J_{\min}={\delta ^{ - 1}}$
			
			     ${{L_{{\rm{opt}}}}}=1$

     \For {$l = 1, \ldots ,{L_{\max }}$}  {
		
		      ${{e_{(l)}}[n]}={ {d_{(l)}}[n] - {\bf{\hat h}}_{(l)}^H[n - 1]{\bf{x}}[n]}$

					${{{{\bf{\hat h}}}_{(l)}}[n]}={  {{{\bf{\hat h}}}_{(l)}}[n - 1] + e_{(l)}^*[n]{{\bf{k}}}[n]}$

	        ${d_{(l + 1)}}[n] = {e_{(l)}}[n]T[n]$

				  ${\pi _{(l + 1)}}[n] = (1 - z){\pi _{(l + 1)}}[n - 1] + z{\left| {{d_{(l + 1)}}[n]} \right|^2}$

				  $J={\pi _{(l + 1)}}[n] - r(l)$

				  \If {$ J < J_{\min}$ }{
					
					    $J_{\min}=J$
            
              ${{L_{{\rm{opt}}}}}=l$

        }				
	
		 }
      
		 ${\bf{\tilde h}}[n] = \sum\limits_{l = 1}^{{L_{{\rm{opt}}}}} {{{{\bf{\hat h}}}_{(l)}}[n]} $
    
     }
\end{algorithm}

Speaking about the computational complexity, the classic implementation of RLS algorithm  requires $4M^2+3M+1$ multiplications, $4M^2-M$ additions, and $M$ divisions for each sample time. In this regard, Table 1 represents the complexity of different steps of m-RLS algorithm in Algorithm 1 by using the classic implementation of each RLS. As shown in this table, the total complexity of m-RLS is of order ${\cal O}({M^2})$ and higher than that of RLS.

It is worth mentioning that we can use low-complexity versions of RLS to implement m-RLS. For instance, by using the transversal DCD algorithm \cite{R30}, the RLS complexity can be reduced to only $3M$ multiplications and $2M{N_{{\rm{itr}}}}+6M$ additions, where ${N_{{\rm{itr}}}}$ is the number of iterations of the DCD algorithm. This approach can be used to implement m-RLS for reducing the complexity of steps 3, 4, 9, and 10. In addition, when  $z = {2^{ - a}}$ and $a$ is an integer, the multiplication by $z$ in step 12 is only a shift in a bit register. Table 2 shows the complexity of m-RLS by using the transversal DCD algorithm. Accordingly, the m-RLS complexity in Table 2 is of order ${\cal O}({M})$ and significantly lower than that of the classic implementation in Table 1.
\begin{table}[t]
\begin{center}
\vspace{5pt}
\caption{Table 1. Complexity of the proposed estimator in Algorithm 1, by using  classic implementation.}
\vspace{5pt}
\begin{tabular}{ |c|c|c|c| }
 \hline
 Step&Mult&Add&Div\\
 \hline
 3&$2M^2+M$&$2M^2-M$&$M$\\
 4&$4M^2+3M+1$&$4M^2-M$&-\\
 5&$M$&$M$&-\\
 9&${L_{\max }}M$&${L_{\max }}M$&-\\
 10&${L_{\max }}M$&${L_{\max }}M$&-\\
 11&${L_{\max }}$&-&-\\
 12&$4{L_{\max }}$&${L_{\max }}$&-\\
 13& -  &${L_{\text{max} }}$&-\\
 14-16& -  &${L_{\text{max} }}$&-\\
 17&-&$({L_{\text{opt} }}-1)M$&-\\
 \hline
 \multicolumn{4}{|l|}{Total: $6M^2+(2{L_{\max }}+5)M+5{L_{\max }}+1$ Mult} \\
\multicolumn{4}{|l|}{ $6M^2+(2{L_{\max }}+{L_{\text{opt} }}-2)M+3{L_{\max }}$ Add} \\
 \multicolumn{4}{|l|}{$M$ Div} \\
 \hline
\end{tabular}
\end{center}
\end{table}

\begin{table}[t]
\begin{center}
\vspace{5pt}
\caption{Table 2. Complexity of the proposed estimator in Algorithm 1, by using transversal DCD implementation.}
\vspace{5pt}
\begin{tabular}{ |c|c|c|c| }
 \hline
 Step&Mult&Add\\
 \hline
 3-4&$M$&$2M$\\
 5&$M$&$M$\\
 9&${L_{\max }}M$&${L_{\max }}M$\\
 10&${L_{\max }}M$&${L_{\max }}(3M+2M{N_{{\rm{itr}}}})$\\
 11&${L_{\max }}$&-\\
 12&${L_{\max }}$&${L_{\max }}$\\
 13& -  &${L_{{\max} }}$\\
 14-16& -  &${L_{{\max} }}$\\
 17&-&$({L_{\text{opt} }}-1)M$\\
 \hline
 \multicolumn{3}{|l|}{Total: $(2{L_{\max }}+2)M+{L_{\max }}$ Mult} \\
\multicolumn{3}{|l|}{$([2{N_{{\rm{itr}}}}+4]{L_{\max }}+{L_{\text{opt} }}+2)M+3{L_{\max }}$ Add} \\
 \hline
\end{tabular}
\end{center}
\end{table}

\section{Simulation Results}
\begin{figure}[bt!]
\centering
  \includegraphics[width=2.5in]{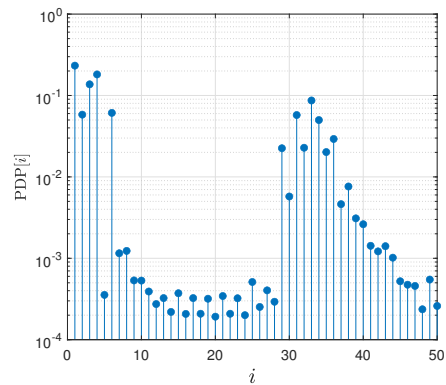}
  \caption{The PDP of the unknown system's IR \cite{R33}.}
\vspace{-15pt}
	\label{F2}
\end{figure}

Digital self-interference cancellation in a full-duplex communication is one of the applications which requires accurate real-time tracking of the self-interference channel \cite{R32}. It this regard, for numerical evaluations, we consider that the unknown system is a self-interference channel, with length $M=50$ and the power-delay-profile (PDP) shown in Fig.~\ref{F2}. The input signal is an uncorrelated BPSK sequence with length $3000$. The forgetting factor in the RLS and m-RLS algorithms is set to $\lambda  = 1 - \frac{1}{{2M}}$. In m-RLS, ${L_{{\rm{max}}}}=5$, and $z=2^{-5}$. The reported results are the averages on $2000$ rounds of simulations.
  
For a better explanation, we split the simulations into three parts. In the first part, the performance of m-RLS is assessed in a fixed coherence length and SNR. In the second part, we compare the simulation results for different coherence lengths and SNRs. Finally, in the third part, we evaluate the performances in a situation where the IR impulsively changes.

\subsection{Part One}
\begin{figure}[bt!]
\centering
  \includegraphics[width=3.2in]{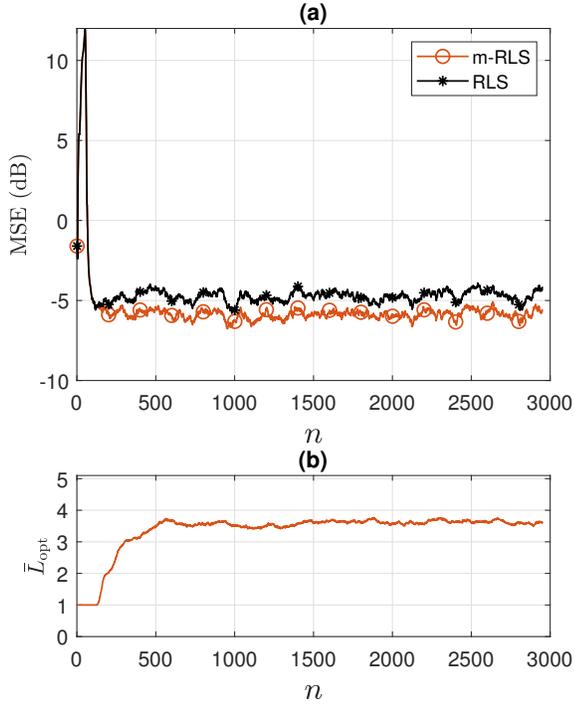}
  \caption{When $N=200$ and SNR is 20 dB, (a) MSE performances of m-RLS and RLS, (b) the average of the optimum number of layers in m-RLS.}
\vspace{-10pt}
	\label{F3}
\end{figure}

In the first part of the simulations, we perform the simulation for the case where the coherence length of the IR taps is $N=200$, and SNR is $20$ dB. Fig.~\ref{F3}-a compares the IR estimation MSE results for m-RLS and RLS; and Fig.~\ref{F3}-b shows the average of the optimal number of layers, ${{\bar L}_{{\rm{opt}}}}$, in m-RLS. As seen, after the initial transience, the MSE of m-RLS is around $1.5$ dB lower than that of RLS. This result shows that how the multi-layered idea can reduce the error of tracking a time-varying system. Based on Fig.~\ref{F3}-b, this supremacy is achieved by using ${{\bar L}_{{\rm{opt}}}} \approx 3.5$.

\begin{figure}[bt!]
\centering
  \includegraphics[width=3.5in]{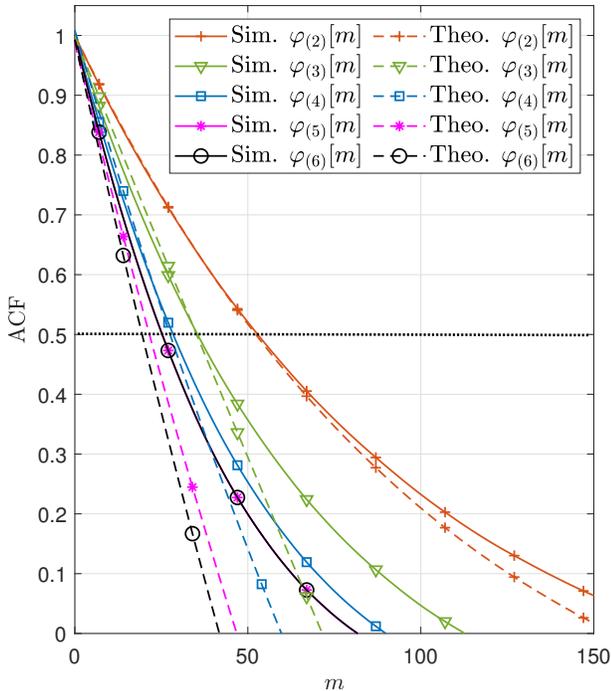}
  \caption{The numerical and theoretical ACF of effective IRs in m-RLS, when $N=200$.}
	\vspace{-10pt}
	\label{F5}
\end{figure}

To confirm our derivation about the ACFs of the effective IRs in m-RLS, Fig.~\ref{F5} compares the theoretical ${\varphi _{(l + 1)}}[m]$ in \eqref{17_5} with that inferred from the numerical simulations.  Based on this figure, the theoretical ACFs roughly match the numerical results for ${\varphi _{(l + 1)}}[m] \ge 0.5$. The slight mismatch, for $l=4 ,5$, comes from the assumptions that we hold to simplify our derivation in Section III-C. It should be mentioned that, according to Fig.~\ref{F5}, the ACF of the effective IR at each layer is narrower than that of the previous layer bringing up this fact that each effective IR varies faster than that in the previous layer.

Furthermore, based on the plots in Fig.~\ref{F5}, we compare the numerical and theoretical coherence lengths ${N_{(l+ 1)}}$ in Table 3. It is noteworthy that the numerical and theoretical coherence lengths are mutually close and justify the derivations in Section III-C.  

\begin{table}[b!]
\begin{center}
\caption{Table 3. The numerical and theoretical coherence length of effective IRs, when $N=200$.}
\vspace{5pt}
\begin{tabular}{ |c|c|c|c| }
 \hline
 $l$&Sim. ${N _{(l + 1)}}$ &Theo. ${N _{(l + 1)}}$\\
 \hline
 1&$53$&$53$\\
 2&$33$&$33$\\
 3&$26$&$26$\\
 4&$23$&$22$\\
 5&$23$&$19$\\
 \hline
\end{tabular}
\end{center}
\end{table}

\subsection{Part Two}
In the second part of the simulations, we inquire about the performance of m-RLS in different SNRs when the time-variation of the system is relatively fast, and slow with the coherence length $N=200$, and $2000$, respectively. 
In addition to the classic RLS, we also compare m-RLS performance with two types of RLS-VFF algorithms. We title these algorithms as RLS-VFF1 and RLS-VFF2 which are proposed in \cite{R5} and \cite{R3}, respectively. The required parameters for RLS-VFF1 and RLS-VFF2 are set based on the original works.

\begin{figure}[bt!]
\centering
  \includegraphics[width=3.2in]{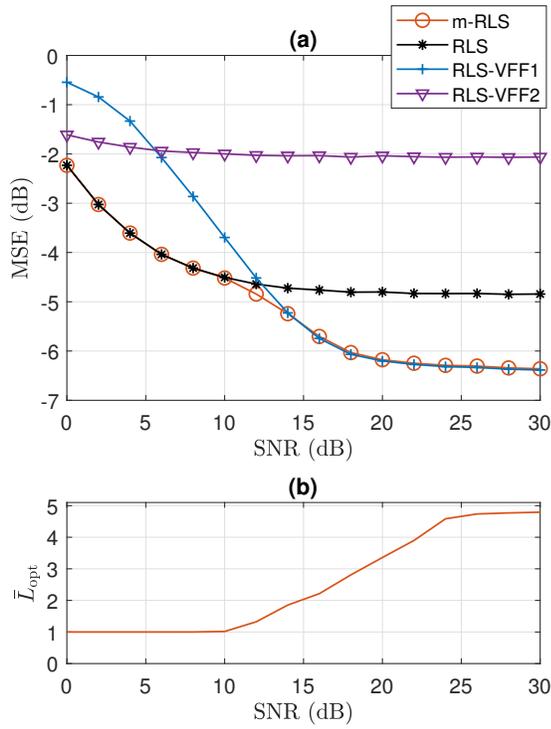}
  \caption{When the input signal is BPSK and $N=200$, (a) the MSE performances for different SNRs, (b) the average of the optimum number of layers in m-RLS for different SNRs.}
\vspace{-10pt}
	\label{F6}
\end{figure}

Fig.~\ref{F6}-a shows the MSE results for $N=200$. Fig.~\ref{F6}-b shows the corresponding ${{\bar L}_{{\rm{opt}}}}$ in m-RLS. As seen, when the system is relatively fast time-varying, for the SNRs less than 10 dB, ${{\bar L}_{{\rm{opt}}}}=1$ and m-RLS is equivalent to RLS. However, for SNRs greater than 10 dB, ${{\bar L}_{{\rm{opt}}}}$ is larger than one, and the performance of m-RLS becomes better than that of RLS. As the SNR increases,  ${{\bar L}_{{\rm{opt}}}}$ grows larger and the MSE performance of m-RLS becomes better than that of RLS. These results establish the fact that a higher SNR leads to a larger ${L_{{\rm{opt}}}}$. The MSE of RLS-VFF1, for SNRs below 14 dB, is worst than that of m-RLS; however, for SNRs above 14 dB, merges to the MSE of m-RLS. The performance of RLS-VFF2 does not remarkably change with the SNR. This estimator has the highest MSE amongst the tested methods, except for very low SNRs, where it outperforms RLS-VFF2.

\begin{figure}[bt!]
\centering
  \includegraphics[width=3.2in]{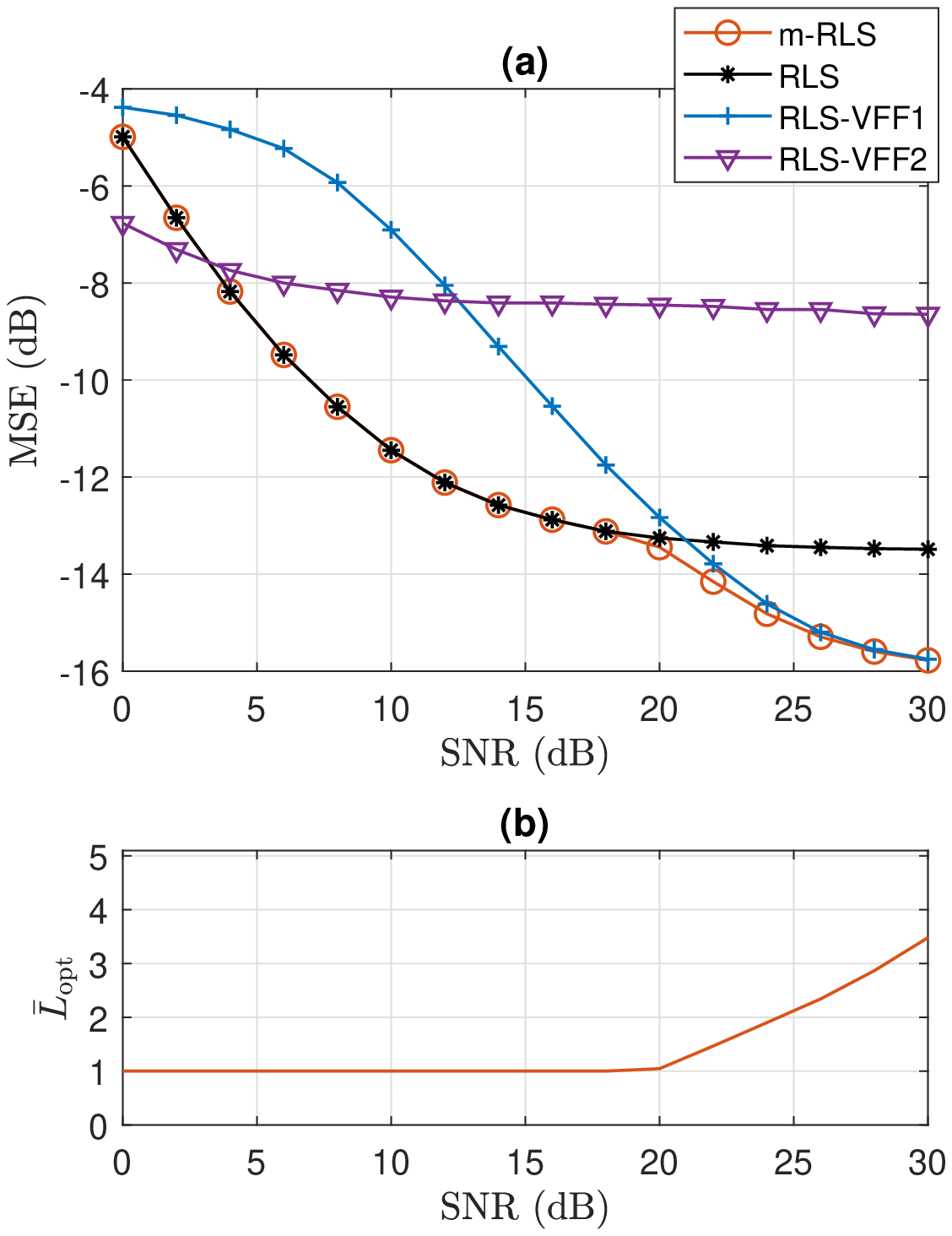}
  \caption{When the input signal is BPSK and $N=2000$, (a) the MSE performances for different SNRs, (b) the average of the optimum number of layers in m-RLS for different SNRs.}
	\vspace{-10pt}
	\label{F7}
\end{figure}

Fig.~\ref{F7} shows the results when $N=2000$. As seen, when the system is relatively slow time-varying, for SNR equal to $20$ dB and higher, ${{\bar L}_{{\rm{opt}}}}$  becomes larger than one and m-RLS outperforms RLS. Comparing Fig.~\ref{F7}-b to Fig.~\ref{F6}-b, one can see that ${{\bar L}_{{\rm{opt}}}}$  is generally larger for the relatively faster time-varying IR.

\begin{figure}[bt!]
\centering
  \includegraphics[width=3.2in]{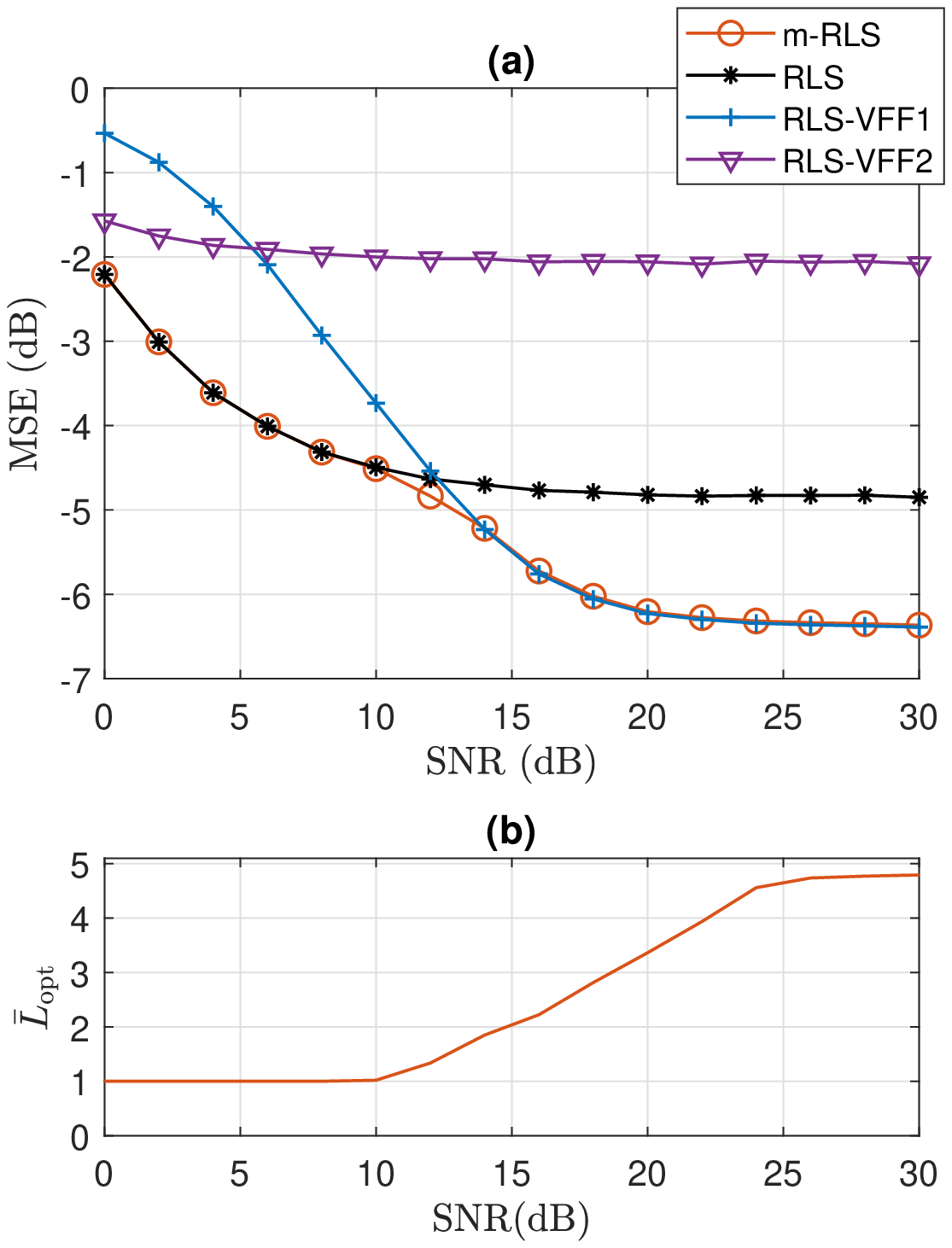}
  \caption{When the input signal is AWGN and $N=200$, (a) the MSE performances for different SNRs, (b) the average of the optimum number of layers in m-RLS for different SNRs.}
	\vspace{-15pt}
	\label{F8}
\end{figure}

 We repeat the same simulation as that in Fig.~\ref{F6} for the case where the input signal is a normalized complex-valued AWGN sequence. The results are presented in Fig.~\ref{F8}, which are fairly identical to those in Fig.~\ref{F6}. Thus, although our formulations in this paper are for BPSK input signal, the derivations and results can also be extended to the case where the input signal is AWGN.

\begin{figure}[bt!]
\centering
  \includegraphics[width=3.2in]{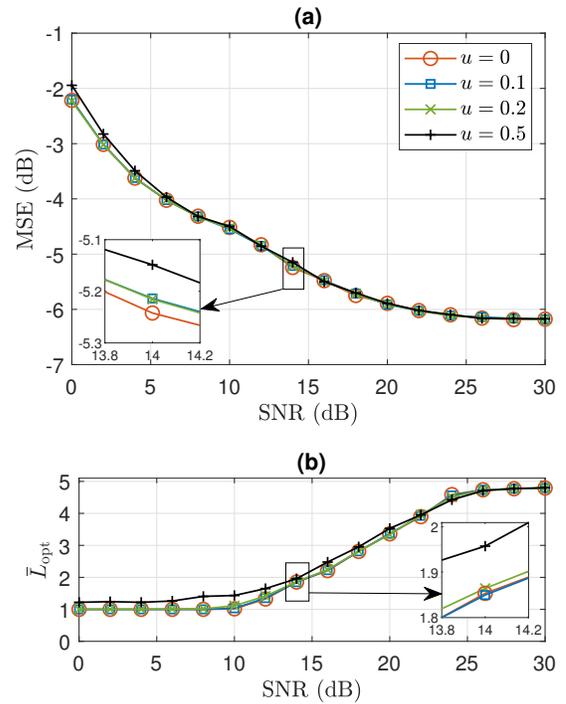}
  \caption{When the input signal is BPSK, $N=200$, and the noise power is uncertain in the m-RLS algorithm as $(1 + u \times randn)\sigma _w^2$, (a) the MSE performances of m-RLS for different SNRs, (b) the average of the optimum number of layers in m-RLS for different SNRs.}
	\vspace{-15pt}
	\label{F9}
\end{figure}

It is worth noting that, for determining the optimal number of layers based on \eqref{166_288}, the noise power must be known in the m-RLS algorithm. However, via numerical evaluations, we show that a moderate uncertainty about the noise power does not significantly affect the performance. To this purpose, Fig.~\ref{F9} shows the performance of m-RLS with the same simulation parameters used in Fig.~\ref{F6}, except it is assumed the uncertain noise power is given to the algorithm as $(1 + u \times randn)\sigma _w^2$, where $randn$ is a zero-mean normalized Gaussian random variable and $u$ is a parameter to control the extension of the uncertainty. As it can be seen from Fig.~\ref{F9}-a and Fig.~\ref{F9}-b, for $u=0.1$ and $0.2$, the MSE performance and ${{\bar L}_{{\rm{opt}}}}$ are the same as those for $u=0$ (no uncertainty). On the other hand, for $u=0.5$, the MSE performance for SNRs lower than $14$ dB becomes slightly worst than that of $u=0$. In this SNR range, ${{\bar L}_{{\rm{opt}}}}$ is higher than one. These results indicate that even up to $50\%$ uncertainty on the noise power does not significantly affect the performance of m-RLS. 

\subsection{Part Three}

\begin{figure}[bt!]
\centering
  \includegraphics[width=3.2in]{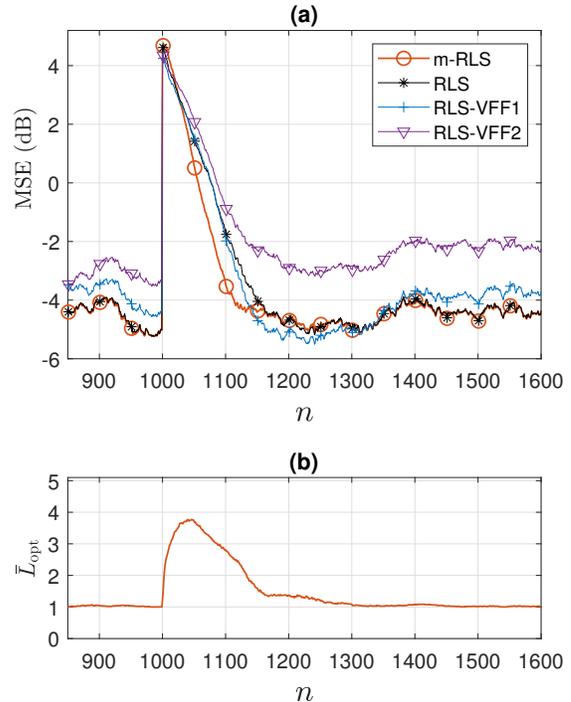}
  \caption{When the input signal is BPSK, $N=200$, SNR=10 dB, and the impulsive change occurs at $n=1000$, a) the MSE performances, (b) the average of the optimum number of layers in m-RLS.}
\vspace{-15pt}
	\label{F10}
\end{figure}

\begin{figure}[bt!]
\centering
  \includegraphics[width=3.2in]{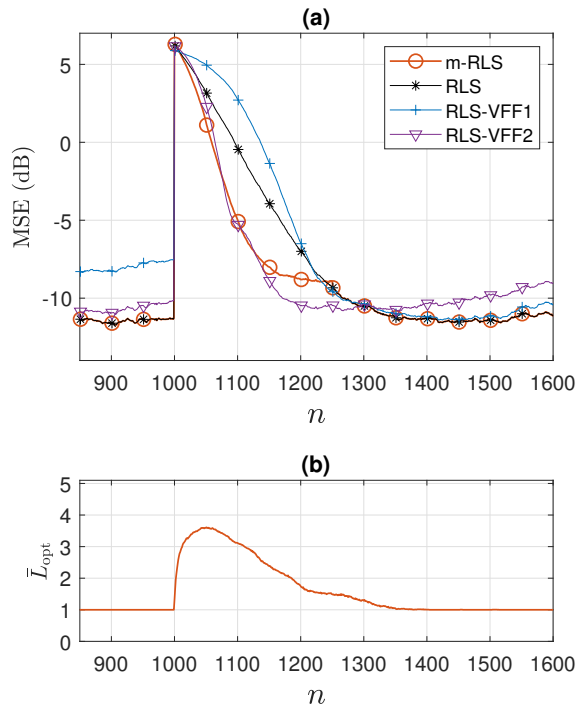}
  \caption{When the input signal is BPSK, $N=2000$, SNR=10 dB, and the impulsive change occurs at $n=1000$, a) the MSE performances, (b) the average of the optimum number of layers in m-RLS.}
	\vspace{-15pt}
	\label{F11}
\end{figure}

In this part, we investigate how an impulsive change, besides the continuous time-variations of the system, is handled in m-RLS. To simulate this scenario, we consider that the IR is time-varying with the coherence time $N$; in addition to this continuous variation, a significant impulsive change is also  imposed to the IR at time $n=1000$ by multiplying all taps of the IR vector by $-1$. When SNR is $10$ dB, Fig.~\ref{F10} and Fig.~\ref{F11} show the results for $N=200$ and $2000$, respectively. 

According to Fig.~\ref{F10}-a, when $N=200$ and the IR is relatively fast time-varying, the MSE levels of m-RLS and RLS are almost the same, before the impulsive change at $n=1000$. Within this period, the MSE of RLS-VFF1 is slightly higher than that of m-RLS and the MSE of RLS-VFF2 has the worst performance confirming the achievement observed in Fig.~\ref{F6}-a at SNR 10 dB. Nevertheless, once the impulsive change occurs, m-RLS has the fastest convergence in tracking this change. The MSE of RLS-VFF1, RLS, and RLS-VFF2 drop slower, respectively. After the transience interval (caused by the impulsive change), the MSE levels turn back to the same order as it that before the impulsive change.
 
It is interesting to notice that, based on Fig.~\ref{F10}-b, ${{\bar L}_{{\rm{opt}}}}  \approx 1$   before the impulsive change. Once the IR is impulsively changed, ${{\bar L}_{{\rm{opt}}}}$ raises up $3.7$ and leads to the supremacy of m-RLS versus the other methods. Within the transience interval, the average of the optimum number of layers settles down to ${{\bar L}_{{\rm{opt}}}}\approx 1$ again. 

Fig.~\ref{F11} shows the results when the IR is relatively slow time-varying with $N=2000$. Based on Fig.~\ref{F11}-a, before and after the transient interval, m-RLS and RLS have the lowest MSE. The RLS-VFF2 and RLS-VFF1 methods have respectively the higher MSE levels verifying the results of Fig.\ref{F7}-a at SNR $10$ dB. Once the impulsive change happens, RLS-VFF2 and m-RLS attain the fastest convergence. RLS and RLS-VFF1 show slower drops, respectively.  
According to Fig.~\ref{F11}-b, ${{\bar L}_{{\rm{opt}}}} $ jumps up to $3.6$ once the impulsive change takes place and settles down to  ${{\bar L}_{{\rm{opt}}}} =1$ when the transience interval ends. 

Based on Fig.~\ref{F10}-b and Fig.~\ref{F11}-b, the optimum number of layers in m-RLS not only depends on the coherence length of the continuously time-variation of the IR but also can be affected by impulsive changes.

\section{Conclusions}
In this paper, we propose the m-RLS adaptive filtering to enhance the accuracy of tracking rapidly time-varying systems. We show that in m-RLS, the power of the lag error and the noise effect are functions of the number of layers. We provide a method to determine the optimum number of layers to minimize the sum of the lag error and the noise effect. The coherence length of the effective IRs in m-RLS are evaluated and it is derived that each effective IR varies faster than that in the previous layer. The implementation complexity of m-RLS is studied and it is shown that  m-RLS is more complex than RLS; however, the complexity order of the proposed approach can be reduced to the order of ${\cal O}(M)$, where $M$ indicates the IR length. 

Based on the simulation results, we attain that, for either a relatively faster time-varying system or a higher SNR, the optimum number layers becomes larger. With the optimum number of layers, m-RLS outperforms the classic RLS and the investigated RLS methods with a variable forgetting factor especially when the system rapidly time-varying. In addition, it is demonstrated that an uncertain knowledge about the noise power does not significantly deteriorate the performance of the proposed method.
%\vspace{-0in}
\begin{appendices}
\section{Proof of \eqref{77_5}}
We split this proof into three sub-proofs:
\subsection{Proof of $\mathbb{E}{|| {{\bf{B}}[n]({\bf{h}}[n] - {\bf{\hat h}}[n - N])} ||^2}={\rho ^N}\mathbb{E}|| {\bf{h}}[n] - $ ${\bf{\hat h}}[n - N] ||^2$}

Let us define ${\bf{a}} = {\bf{h}}[n] - {\bf{\hat h}}[n - N]$. According to the definition of ${\bf{\Theta }}[n]$, one can see that 
\begin{align}
\mathbb{E}{\left\| {{\bf{\Theta }}[n]{\bf{a}}} \right\|^2} =& \sum\limits_{i = 0}^{M - 1} {[\mathbb{E}{{\left| {{a_i}} \right|}^2} - 2\varepsilon {\mathop{\rm Re}\nolimits} \sum\limits_{j = 0}^{M - 1} {\mathbb{E}\left\{ {{x_i}[n]x_j^*[n]a_i^*{a_j}} \right\}} }  \nonumber\\
&+{\varepsilon ^2}\mathbb{E}{\left| {\sum\limits_{j = 0}^{M - 1} {{x_i}[n]} x_j^*[n]{a_i}} \right|^2}],
\label{appB_1}
\end{align}
where ${x_i}[n]$ and $a_i$ are the $i$th entries of ${{\bf{x}}[n]}$ and ${\bf{a}}$, respectively. Since ${x_i}[n]$'s and $a_i$'s  are uncorrelated zero-mean variables, $\mathbb{E}\left\{ {{x_i}[n]x_j^*[n]a_i^*{a_j}} \right\}={\delta _{i - j}}{\mathbb{E}{{\left| {{a_i}} \right|}^2}}$, and $\mathbb{E}{| {\sum\limits_{j = 0}^{M - 1} {{x_i}[n]} x_j^*[n]{a_i}} |^2}=M {\mathbb{E}{{\left| {{a_i}} \right|}^2}}$. As a result, 
\begin{align}
\mathbb{E}{\left\| {{\bf{\Theta }}[n]{\bf{a}}} \right\|^2}= \rho\sum\limits_{i = 0}^{M-1} {\mathbb{E}{{{\left| {{a_i}} \right|}^2}}}=\rho \mathbb{E}{\left\| {{\bf{a}}} \right\|^2},
\label{appB_2}
\end{align}
in which $\rho=1 - 2\varepsilon  + {\varepsilon ^2}M$. 
 In the next step, we define ${\bf{b}} = {\bf{\Theta }}[n]{\bf{a}}$. In the same way, since ${\bf{\Theta }}[n-1]$ and ${\bf{b}}$ are uncorrelated, $\mathbb{E} {\left\| {{\bf{\Theta }}[n - 1]{\bf{b}}} \right\|^2} = \rho \mathbb{E}{\left\| {\bf{b}} \right\|^2} = {\rho^2}{\left\| {\bf{a}} \right\|^2}$. Continuing this procedure and recalling the definition of ${\bf{B}}[n]$ in \eqref{7_2}, it is concluded that
\begin{align}
\mathbb{E}{\left\| {{\bf{B}}[n]{\bf{a}}} \right\|^2}&=\mathbb{E}{\left\| {\prod\limits_{k = 0}^{N - 1} {{\bf{\Theta }}[n - k]}{\bf{a}}} \right\|^2}= \rho^N \mathbb{E}{\left\| {{\bf{a}}} \right\|^2} \nonumber\\
&=\rho^N \mathbb{E}{\left\| {{\bf{h}}[n] - {\bf{\hat h}}[n - N]} \right\|^2}.
\label{appB_3}
\end{align}

\subsection{Proof of $\mathbb{E}{|| {\bf{h}}[n] - {\bf{\hat h}}[n - N] ||^2}=\mathbb{E}{\left\| {{{\bf{h}}}[n]} \right\|^2} + {\varepsilon ^2}M\sigma _w^2$}

We can write
\begin{align}
\mathbb{E}{\left\| {{\bf{h}}[n] - {\bf{\hat h}}[n - N]} \right\|^2} =& \mathbb{E}{\left\| {{\bf{h}}[n]} \right\|^2} + \mathbb{E}{\left\| {{\bf{\hat h}}[n - N]} \right\|^2} \nonumber\\
&- 2{\mathop{\rm Re}\nolimits}  \mathbb{E}\{{{\bf{h}}^H}[n]{\bf{\hat h}}[n - N] \}.
\label{appB_4}
\end{align}
From the RLS update procedure in \eqref{2_3}, it is seen that ${{\bf{\hat h}}[n]}={{\bf{h}}[n]}+{\bf w'}[n]$, where ${\bf w'}[n]$ is a zero-mean noise vector with  ${\varepsilon ^2}\sigma _w^2$ as the variance of each entry. Assuming that ${\bf w'}[n]$ is uncorrelated from ${{\bf{ h}}[n]}$, we have
\begin{align}
\mathbb{E}{\left\| {{\bf{\hat h}}[n - N]} \right\|^2}=& \mathbb{E}{\left\| {{\bf{h}}[n]} \right\|^2} + \mathbb{E}{\left\| {{\bf{w'}}[n]} \right\|^2} \nonumber\\
=&\mathbb{E}{\left\| {{\bf{h}}[n]} \right\|^2}+{\varepsilon ^2}M\sigma _w^2,
\label{appB_44}
\end{align}
and  
\begin{align}
\mathbb{E}\{{{\bf{h}}^H}[n]{\bf{\hat h}}[n - N]\} =& \sum\limits_{i = 0}^{M-1} \mathbb{E}\{{h_i^*[n] h_i[n - N]}\}  \nonumber\\
=& \varphi [N]\sum\limits_{i = 0}^{M-1} \mathbb{E}{\left| {{h_i}[n]} \right|^2}  \nonumber\\
=& 0.5\mathbb{E}{\left\| {{\bf{h}}[n]} \right\|^2},
\label{appB_5}
\end{align}
where $h_i[n]$ is the $i$th entry of ${{\bf{h}}[n]}$, and $\varphi [n]$ is the normalized ACF (i.e., ${\varphi [0]}=1$) of $h_i[n]$, for $i = 0, \ldots ,M-1$ (note that it is assumed that all $h_i[n]$'s have the same ACF). In \eqref{appB_5}, we let ${\varphi [N]}=0.5$ due to the fact that the ACF meets $0.5$ at the coherence length \cite{R31}. Then, by considering \eqref{appB_44} and \eqref{appB_5} into \eqref{appB_4}, we have
\begin{align}
\mathbb{E}{|| {\bf{h}}[n] - {\bf{\hat h}}[n - N] ||^2}=\mathbb{E}{\left\| {{{\bf{h}}}[n]} \right\|^2} + {\varepsilon ^2}M\sigma _w^2.
\label{appB_66}
\end{align}

\subsection{Proof of $\mathbb{E}{\left\| {{\bf{c}}[n]} \right\|^2} = {\varepsilon ^2}M\left( {\frac{{1 - {\rho ^N}}}{{1 - \rho }}} \right)\sigma _w^2$}

The definition of ${{\bf{c}}[n]}$ is brought in \eqref{7_2}. Since the noise samples are uncorrelated,
\begin{align}
\mathbb{E}{\left\| {{\bf{ c}}[n]} \right\|^2} = \mathbb{E}{\left\| {{\boldsymbol{\gamma }}[n]} \right\|^2}+{\sum\limits_{k = 1}^{N - 1} {\mathbb{E}\left\| {(\prod\limits_{i = 0}^{k - 1} {{\bf{\Theta }}[n - i]} ){\boldsymbol{\gamma }}[n - k]} \right\|} ^2}.
\label{appB_55}
\end{align}
On the other hand, from the first sub-section in Appendix A, we know that $\mathbb{E}{|| {(\prod\limits_{i = 0}^{k - 1} {{\bf{\Theta }}[n - i]} ){\boldsymbol{\gamma }}[n - k]} ||^2} = {\rho ^k}\mathbb{E}{\left\| {{\boldsymbol{\gamma }}[n - k]} \right\|^2} = {\rho ^k}{\varepsilon ^2}M\sigma _w^2$. Replacing this in \eqref{appB_55}, we conclude that
\begin{align}
\mathbb{E}{\left\| {{\bf{ c}}[n]} \right\|^2} = {\varepsilon ^2}M\sigma _w^2\sum\limits_{k = 0}^{N - 1} {{\rho ^k}} ={\varepsilon ^2}M {\frac{{1  - {\rho ^N}}}{{1 - \rho }}} \sigma _w^2
\label{appB_6}
\end{align}

Finally, considering the results in \eqref{appB_3}, \eqref{appB_66}, and \eqref{appB_6} together leads to \eqref{77_5}.

\section{Proof of \eqref{17_5}}
Since we assume the IRs and their estimates are uncorrelated with the input signal, discarding the noise in \eqref{16_20} leads to 
\begin{align}
\begin{array}{l}
\mathbb{E}\{{\bf{h}}_{(l + 1)}^H[n]{{\bf{h}}_{(l + 1)}}[n - m] \}= \\
\sum\limits_{i = 0}^{M - 1} {\sum\limits_{j = 0}^{M - 1} {\mathbb{E}\left\{ {{Q_{i,j}}} \right\}} } \mathbb{E}\left\{ {{{({h_{(l),i}}[n] - {{\hat h}_{(l),i}}[n - {N_{(l)}}])}^*}} \right.\\
\,\,\,\,\, \,\,\,\,\, \,\,\,\,\, \,\,\,\,\, \,\,\,\,\, \left. {({h_{(l),j}}[n - m] - {{\hat h}_{(l),j}}[n - m - {N_{(l)}}])} \right\},
\end{array}
\label{appc_1}
\end{align}
where ${Q_{i,j}}$ is the $(i,j)$th entry of matrix ${\bf{Q}} = {\bf{B}}_{(l)}^H[n]{{\bf{B}}_{(l)}}[n - m]$.  Also, $h_{(l),i}[n]$ and ${{\hat h}_{(l),i}}[n]$ are the $i$th entry of ${\bf{h}}_{(l)}[n]$ and ${{{\bf{\hat h}}}_{(l)}}[n]$, respectively. Considering that the tap-weights in an effective IR are zero-mean and mutually uncorrelated variables, we can rewrite \eqref{appc_1} as
\begin{align}
\begin{array}{*{20}{l}}
\mathbb{E}{\{ {\bf{h}}_{(l + 1)}^H[n]{{\bf{h}}_{(l + 1)}}[n - m]\}  = }\\
{\sum\limits_{i = 0}^{M - 1} \mathbb{E}{\left\{ {{Q_{i,i}}} \right\}} \mathbb{E}\left\{ {{{({h_{(l),i}}[n] - {{\hat h}_{(l),i}}[n - {N_{(l)}}])}^*}} \right.}\\
{{\kern 1pt} {\kern 1pt} {\kern 1pt} {\kern 1pt} {\kern 1pt} {\kern 1pt} {\kern 1pt} {\kern 1pt} {\kern 1pt} {\kern 1pt} {\kern 1pt} {\kern 1pt} {\kern 1pt} {\kern 1pt} {\kern 1pt} {\kern 1pt} {\kern 1pt} {\kern 1pt} {\kern 1pt} {\kern 1pt} {\kern 1pt} {\kern 1pt} {\kern 1pt} {\kern 1pt} {\kern 1pt} \left. {({h_{(l),i}}[n - m] - {{\hat h}_{(l),i}}[n - m - {N_{(l)}}])} \right\}}.
\end{array}
\label{appc_11}
\end{align}

Based on the definition of ${\bf{B}}_{(l)}[n]$ in \eqref{16_21}, one can realize that, when $m \le {N_{(l)}}$,
\begin{align}
{Q_{i,i}} =& {\prod\limits_{k = 0}^{m - 1} {(1 - \varepsilon {{\left| {{x_i}[n - k]} \right|}^2})} }  \nonumber\\
& {\prod\limits_{k = m}^{{N_{(l)}} - 1} {(1 + ({\varepsilon ^2}M - 2\varepsilon ){{\left| {{x_i}[n - m - k]} \right|}^2})} }  \nonumber\\
& {\prod\limits_{k = {N_{(l)}}}^{m + {N_{(l)}} - 1} {(1 - \varepsilon {{\left| {{x_i}[n - k]} \right|}^2})} }
\label{appc_2}
\end{align}
and, when  $m > {N_{(l)}}$,
\begin{align}
{Q_{i,i}} = &{\prod\limits_{k = 0}^{{N_{(l)}} - 1} {(1 - \varepsilon {{\left| {{x_i}[n - k]} \right|}^2})} }  \nonumber\\
&  {\prod\limits_{k = m}^{m + {N_{(l)}} - 1} {(1 - \varepsilon {{\left| {{x_i}[n - k]} \right|}^2})} }.
\label{appc_3}
\end{align}
Since the input signal $x[n]$ is a BPSK sequence (${{\left| {{x_i}[n]} \right|}^2}=1$), by using \eqref{appc_2} and \eqref{appc_3}, we have 
\begin{align}
{Q_{i,i}} = \left\{ \begin{array}{l}
{\lambda ^{2m}}{\rho ^{{N_{(l)}} - m}}\,;\,\,\,\,{\rm{for}}\,\,\,0 \le m \le {N_{(l)}}\\
{\lambda ^{{2N_{(l)}}}};\,\,\,\,{\rm{for}}\,\,\, {N_{(l)}}<m.
\end{array} \right.
\label{appc_6}
\end{align}

As seen, ${Q_{i,i}}$ is independent of $n$ and $i$, thus, replacing \eqref{appc_6} in \eqref{appc_11} results in 
\begin{align}
&\mathbb{E}{\bf{h}}_{(l + 1)}^H[n]{{\bf{h}}_{(l + 1)}}[n-m] = \nonumber\\
& \mathbb{E} \left\{ {{{({{\bf{h}}_{(l)}}[n] - {{{\bf{\hat h}}}_{(l)}}[n - {N_{(l)}}])}^H}} \right. \nonumber\\
&\,\,\,\,\,\, \left. {({{\bf{h}}_{(l)}}[n - m] - {{{\bf{\hat h}}}_{(l)}}[n - m - {N_{(l)}}])} \right\}{Q_{i,i}}= \nonumber\\
& \mathbb{E}{\left\| {{{\bf{h}}_{(l + 1)}}[n]} \right\|^2} \nonumber\\
&\left( {2{\varphi _{(l)}}[m] - {\varphi _{(l)}}[m - {N_{(l)}}] - {\varphi _{(l)}}[m + {N_{(l)}}]} \right){Q_{i,i}}.
\label{appc_5}
\end{align}
Then, substituting \eqref{appc_5} and \eqref{16_23} in \eqref{17_2}, when the noise is discarded, results in \eqref{17_5}.

\section{Proof of ${\varphi _{(l + 1)}}[m] \approx \exp (-\beta  m )$}

From \eqref{17_5}, we have ${\varphi _{(l + 1)}}[m] = \left( {{f'}[m] + {f''}[m]} \right){q_{(l)}}[m]$, where ${f'}[m] = {\varphi _{(l)}}[m] - {\varphi _{(l)}}[m - {N_{(l)}}]$, and ${f''}[m] = {\varphi _{(l)}}[m] - {\varphi _{(l)}}[m + {N_{(l)}}]$. Since ${\varphi _{(l)}}[m] = \exp ( - \alpha m )$, and ${\varphi _{(l)}}[{N_{(l)}}] = 0.5$, one can see that ${f'}[m] = 0.5\exp ( - \alpha m)$, and ${f''}[m] = 0.5 - \frac{\alpha }{{\log 2}}m$, for $0 \le m \le {N_{(l)}}$. As a result, we can approximate that ${f'}[m] + {f''}[m] \approx \exp ( - 3\alpha m)$.

On the other hand, from \eqref{17_6}, we have ${q_{(l)}}[m] = \exp ( - g m)$, for $0 \le m \le {N_{(l)}}$, where $g = \log(\frac{\rho }{{{\lambda ^2}}})$. Putting these results together, it is concluded that ${\varphi _{(l + 1)}}[m] \approx \exp ( - \beta m)$, where $\beta  = 3\alpha  + g$.

\section{Proof of \eqref{166_277}}

Since the input signal is BPSK, based on \eqref{12}, we have 
\begin{align}
\mathbb{E}{\left| {{d_{(l + 1)}}[n]} \right|^2} =& \mathbb{E}{\left\| {{{\bf{h}}_{(l + 1)}}[n]} \right\|^2} + \sigma _w^2 \nonumber\\ &+ 2{\mathop{\rm Re}\nolimits} \mathbb{E}\{ {\bf{h}}_{(l + 1)}^H[n]{\bf{x}}[n]{w^*}[n]\}.
\label{appE_0}
\end{align}
Let us first investigate $\mathbb{E}\{ {\bf{h}}_{(l + 1)}^H[n]{\bf{x}}[n]{w^*}[n]\}$ in \eqref{appE_0}.
Similar to \eqref{5}, for $l$th layer, we can write
\begin{align}
{{{\bf{\hat h}}}_{(l)}}[n] = {{\bf{h}}_{(l)}}[n] - {\bf{\Theta }}[n]\left( {{{\bf{h}}_{(l)}}[n] - {{{\bf{\hat h}}}_{(l)}}[n - 1]} \right) + {\boldsymbol{\gamma }}[n].
\label{appE_1}
\end{align}
Replacing \eqref{12_add} in \eqref{appE_1} results in
\begin{align}
{{\bf{h}}_{(l + 1)}}[n] = {\bf{\Theta }}[n]\left( {{{\bf{h}}_{(l)}}[n] - {{{\bf{\hat h}}}_{(l)}}[n - 1]} \right) + {\boldsymbol{\gamma }}[n].
\label{appE_2}
\end{align}
By using \eqref{appE_2}, one can expand ${{\bf{h}}_{(l + 1)}}[n]$ as
\begin{align}
{{\bf{h}}_{(l + 1)}}[n] = &{{\bf{\Theta }}^l}[n]{{\bf{h}}_{(1)}}[n] - \sum\limits_{i = 1}^l {{{\bf{\Theta }}^i}[n]{{\bf{\hat h}}_{(l - i + 1)}}[n - 1]} \nonumber\\
 &- \sum\limits_{i = 1}^l {{{\bf{\Theta }}^{i + 1}}[n]{\boldsymbol{\gamma }}[n]}.  
\label{appE_3}
\end{align}
On the right-hand side of \eqref{appE_3}, the first and the second terms are uncorrelated with $w[n]$. The only correlated part is the third term because ${\boldsymbol{\gamma }}[n]=\varepsilon {\bf{x}}[n]{w^*}[n]$. Thus, since ${\bf{\Theta }}[n]$ is a Hermitian matrix, we have
\begin{align}
\mathbb{E}\{ {\bf{h}}_{(l + 1)}^H[n]{\bf{x}}[n]{w^*}[n]\}  =  - \varepsilon \sigma _w^2\mathbb{E}\{ {{\bf{x}}^H}[n]\sum\limits_{i = 1}^l {{{\bf{\Theta }}^{i + 1}}[n]{\bf{x}}[n]} \}.   
\label{appE_4}
\end{align}
Based on the definition of ${{{\bf{\Theta }}}[n]}$ and using the fact that the input is BPSK (i.e., ${{\bf{x}}^H}[n]{\bf{x}}[n] = M$), it is easy to follow that ${{\bf{x}}^H}[n]{{\bf{\Theta }}^{i + 1}}[n]{\bf{x}}[n] = M{(1 - \varepsilon M)^{i + 1}} $ is independent of $n$. As a result, 
\begin{align}
\mathbb{E}\{ {\bf{h}}_{(l + 1)}^H[n]{\bf{x}}[n]{w^*}[n]\}  =  - (1 - {(1 - \varepsilon M)^l})\sigma _w^2.   
\label{appE_4}
\end{align}
Replacing \eqref{appE_4} in \eqref{appE_0} results in \eqref{166_277}.

\end{appendices}

%\vspace{-7pt}
\bibliographystyle{IEEEtran}
\bibliography{References}

\begin{thebibliography}{10}
\providecommand{\url}[1]{#1}
\csname url@samestyle\endcsname
\providecommand{\newblock}{\relax}
\providecommand{\bibinfo}[2]{#2}
\providecommand{\BIBentrySTDinterwordspacing}{\spaceskip=0pt\relax}
\providecommand{\BIBentryALTinterwordstretchfactor}{4}
\providecommand{\BIBentryALTinterwordspacing}{\spaceskip=\fontdimen2\font plus
\BIBentryALTinterwordstretchfactor\fontdimen3\font minus
  \fontdimen4\font\relax}
\providecommand{\BIBforeignlanguage}[2]{{%
\expandafter\ifx\csname l@#1\endcsname\relax
\typeout{** WARNING: IEEEtran.bst: No hyphenation pattern has been}%
\typeout{** loaded for the language `#1'. Using the pattern for}%
\typeout{** the default language instead.}%
\else
\language=\csname l@#1\endcsname
\fi
#2}}
\providecommand{\BIBdecl}{\relax}
\BIBdecl

\bibitem{R2}
B.~Farhang-Boroujeny, \emph{{Adaptive Filters: Theory and Applications}}.\hskip
  1em plus 0.5em minus 0.4em\relax Chichester, U.K.: Wiley, 1998.

\bibitem{R7}
S.~Haykin, \emph{{Adaptive Filter Theory}}.\hskip 1em plus 0.5em minus
  0.4em\relax 4th ed: Prentice-Hall, 2002.

\bibitem{R11}
D.~T.~M. Slock and T.~Kailath, ``{Numerically stable fast transversal filters
  for recursive least-squares adaptive filtering},'' in \emph{{Numerical Linear
  Algebra, Digital Signal Processing and Parallel Algorithms}}, G.~H. Golub and
  P.~Van~Dooren, Eds.\hskip 1em plus 0.5em minus 0.4em\relax Berlin,
  Heidelberg: Springer Berlin Heidelberg, 1991, pp. 605--615.

\bibitem{R14}
J.~Cioffi and T.~Kailath, ``{Fast, recursive-least-squares transversal filters
  for adaptive filtering},'' \emph{IEEE Transactions on Acoustics, Speech, and
  Signal Processing}, vol.~32, no.~2, pp. 304--337, 1984.

\bibitem{R15}
J.-L. Botto, ``{Stabilization of fast recursive least-squares transversal
  filters for adaptive filtering},'' in \emph{ICASSP '87. IEEE International
  Conference on Acoustics, Speech, and Signal Processing}, vol.~12, 1987, pp.
  403--406.

\bibitem{R6}
D.~{Lee}, M.~{Morf}, and B.~{Friedlander}, ``{Recursive least squares ladder
  estimation algorithms},'' \emph{IEEE Transactions on Acoustics, Speech, and
  Signal Processing}, vol.~29, no.~3, pp. 627--641, 1981.

\bibitem{R16}
B.~Friedlander, ``{Lattice filters for adaptive processing},''
  \emph{Proceedings of the IEEE}, vol.~70, no.~8, pp. 829--867, 1982.

\bibitem{R20}
M.~Z.~A. Bhotto and A.~Antoniou, ``{Robust recursive least-squares
  adaptive-filtering algorithm for impulsive-noise environments},'' \emph{IEEE
  Signal Processing Letters}, vol.~18, no.~3, pp. 185--188, 2011.

\bibitem{R21}
S.~Farahmand and G.~B. Giannakis, ``{Robust RLS in the presence of correlated
  noise using outlier sparsity},'' \emph{IEEE Transactions on Signal
  Processing}, vol.~60, no.~6, pp. 3308--3313, 2012.

\bibitem{R22}
S.-C. Chan and Y.-X. Zou, ``{A recursive least M-estimate algorithm for robust
  adaptive filtering in impulsive noise: fast algorithm and convergence
  performance analysis},'' \emph{IEEE Transactions on Signal Processing},
  vol.~52, no.~4, pp. 975--991, 2004.

\bibitem{R23}
Y.~Zou, S.~Chan, and T.~Ng, ``{A recursive least M-estimate (RLM) adaptive
  filter for robust filtering in impulse noise},'' \emph{IEEE Signal Processing
  Letters}, vol.~7, no.~11, pp. 324--326, 2000.

\bibitem{R9}
E.~Eleftheriou and D.~Falconer, ``{Tracking properties and steady-state
  performance of RLS adaptive filter algorithms},'' \emph{IEEE Transactions on
  Acoustics, Speech, and Signal Processing}, vol.~34, no.~5, pp. 1097--1110,
  1986.

\bibitem{R10}
A.~H. Sayed, \emph{{Fundamentals of Adaptive Filtering}}.\hskip 1em plus 0.5em
  minus 0.4em\relax John Wiley \& Sons, 2003.

\bibitem{R3}
C.~{Paleologu}, J.~{Benesty}, and S.~{Ciochina}, ``{A robust variable
  forgetting factor recursive least-squares algorithm for system
  identification},'' \emph{IEEE Signal Processing Letters}, vol.~15, pp.
  597--600, 2008.

\bibitem{R5}
M.~Z.~A. {Bhotto} and A.~{Antoniou}, ``{New improved recursive least-squares
  adaptive-filtering algorithms},'' \emph{IEEE Transactions on Circuits and
  Systems I: Regular Papers}, vol.~60, no.~6, pp. 1548--1558, 2013.

\bibitem{R26}
S.-H. Leung and C.~So, ``{Gradient-based variable forgetting factor RLS
  algorithm in time-varying environments},'' \emph{IEEE Transactions on Signal
  Processing}, vol.~53, no.~8, pp. 3141--3150, 2005.

\bibitem{R28}
B.~Toplis and S.~Pasupathy, ``{Tracking improvements in fast RLS algorithms
  using a variable forgetting factor},'' \emph{IEEE Transactions on Acoustics,
  Speech, and Signal Processing}, vol.~36, no.~2, pp. 206--227, 1988.

\bibitem{R30}
Y.~V. Zakharov, G.~P. White, and J.~Liu, ``{Low-complexity RLS algorithms using
  dichotomous coordinate descent iterations},'' \emph{IEEE Transactions on
  Signal Processing}, vol.~56, no.~7, pp. 3150--3161, 2008.

\bibitem{R13}
E.~Eweda, ``{Comparison of RLS, LMS, and sign algorithms for tracking randomly
  time-varying channels},'' \emph{IEEE Transactions on Signal Processing},
  vol.~42, no.~11, pp. 2937--2944, 1994.

\bibitem{R31}
\BIBentryALTinterwordspacing
R.~Steele and L.~Hanzo, \emph{{Mobile Radio Communications: Second and Third
  Generation Cellular and WATM Systems: 2nd}}.\hskip 1em plus 0.5em minus
  0.4em\relax IEEE Press - John Wiley, May 1999, address: Chichester, UK.
  [Online]. Available: \url{https://eprints.soton.ac.uk/251446/}
\BIBentrySTDinterwordspacing

\bibitem{R32}
D.~Wu, C.~Zhang, S.~Gao, and D.~Chen, ``{A digital self-interference
  cancellation method for practical full-duplex radio},'' in \emph{2014 IEEE
  International Conference on Signal Processing, Communications and Computing
  (ICSPCC)}, 2014, pp. 74--79.

\bibitem{R33}
M.~Towliat, Z.~Guo, L.~J. Cimini, X.-G. Xia, and A.~Song, ``{Self-interference
  channel characterization in underwater acoustic in-band full-duplex
  communications using OFDM},'' in \emph{Global Oceans 2020: Singapore – U.S.
  Gulf Coast}, 2020, pp. 1--7.

\end{thebibliography}

\end{document}